\documentclass[aps,prd,12pt]{revtex4}

\usepackage{epsfig} 
\usepackage{bm}
\def\drm{{\rm d}}
\def\xbr{\ensuremath{\sigma B}}

\begin{document}
\title{Anomalously interacting new extra vector bosons and their first LHC constraints}
\author{M. V. Chizhov}
\affiliation{Centre for Space Research and Technologies, Faculty of Physics,
University of Sofia, 1164 Sofia, Bulgaria}
\affiliation{Dzhelepov Laboratory of Nuclear Problems,\\
\mbox{Joint Institute for Nuclear Research, 141980, Dubna, Russia}}

\author{V. A. Bednyakov}
\affiliation{Dzhelepov Laboratory of Nuclear Problems,\\
\mbox{Joint Institute for Nuclear Research, 141980, Dubna, Russia}}

\author{I. R. Boyko}
\affiliation{Dzhelepov Laboratory of Nuclear Problems,\\
\mbox{Joint Institute for Nuclear Research, 141980, Dubna, Russia}}

\author{J. A. Budagov}
\affiliation{Dzhelepov Laboratory of Nuclear Problems,\\
\mbox{Joint Institute for Nuclear Research, 141980, Dubna, Russia}}

\author{M. A. Demichev}
\affiliation{Dzhelepov Laboratory of Nuclear Problems,\\
\mbox{Joint Institute for Nuclear Research, 141980, Dubna, Russia}}

\author{I. V. Yeletskikh}
\affiliation{Dzhelepov Laboratory of Nuclear Problems,\\
\mbox{Joint Institute for Nuclear Research, 141980, Dubna, Russia}}


\date{14.~10.~2011}

\begin{abstract}
     In this review phenomenological consequences of the Standard Model
     extension by means of new spin-1 chiral fields with the internal quantum
     numbers of the electroweak Higgs doublets are summarized.
     The prospects for resonance production and detection
     of the chiral vector $Z^*$ and $W^{*\pm}$ bosons
     at the LHC energies are considered on the basis of
     quantitative simulations   within the CompHEP/CalcHEP package.
     The $Z^*$ boson can be observed as a Breit-Wigner resonance peak
     in the invariant dilepton mass distributions
     in the same way as the well-known extra gauge $Z'$ bosons.
     However, the $Z^*$ bosons have unique signatures in transverse momentum,
     angular and pseudorapidity distributions of the final leptons,
     which allow one to distinguish them from other heavy neutral resonances.

     In 2010, with 40~pb$^{-1}$ of the LHC proton-proton data at the 
     energy 7~TeV, the ATLAS detector was used to search for narrow resonances
     in the invariant mass spectrum 
     of $e^+e^-$ and $\mu^+\mu^-$ final states and high-mass charged states
     decaying to a charged lepton and a neutrino.
     No statistically significant excess above the Standard Model expectation 
     was observed. 
     The exclusion mass limits of 1.15~TeV$/c^2$ and 1.35~TeV$/c^2$
     were obtained for the chiral neutral $Z^*$ and charged $W^*$ bosons, respectively.
     These are the first direct
     limits on the $W^*$ and $Z^*$ boson production.

     Based on the above, a novel strategy for the chiral boson search 
     in the LHC dijet data is discussed.
     For almost all currently considered exotic models the relevant signal 
     is expected in the central dijet rapidity region $y_{1,2}\simeq 0$ and $|y_1-y_2|\simeq 0$.
     On the contrary, the chiral bosons do not contribute to this region
     but produce an excess of dijet events far away from it. 
     In particular, for these bosons the appropriate 
     kinematical restrictions lead to a dip in the centrality ratio 
     distribution over the dijet invariant mass instead of a bump 
     expected in the most exotic  models.
\end{abstract}

\pacs{12.60.-i, 13.85.-t, 14.80.-j}

\maketitle

\section{Introduction} 
      Gauge interactions are the only well-established fundamental
      interactions in Nature.
      The method of the covariant derivatives leads to the unique minimal
      form of the gauge boson couplings to the matter fermions.
      Nevertheless, the Yukawa interactions of the Higgs bosons
      are also necessary for self-consistent construction of the Standard Model (SM).
      Furthermore, although the gauge symmetry allows anomalous interactions in
      the initial Lagrangian, all known fundamental spin-1 bosons
      (photon, $W^\pm$, $Z$ boson and gluons)
      possess only renormalizable minimal interactions with the known fermions.
      The anomalous interactions are considered as effective ones and
      are generated at the level of the quantum loop corrections.
      They are usually proportional to the additional square of a small
      coupling constant and can be neglected in the first-order approximation.

      New heavy neutral gauge bosons are predicted in many extensions of the SM.
      They are associated with additional U(1)$'$ gauge symmetries and
      are generically called $Z'$ bosons.
      The minimal gauge interactions of these bosons with matter lead to
      the well-known angular distribution of outgoing leptons (the $Z'$ decay product)
      in the dilepton center-of-mass reference frame
\begin{equation}\label{sV}{
    \frac{{\rm d}\sigma_{Z'}}{{\rm d}\cos\theta^*}\propto 1+
    A_{\rm FB}\cdot\cos\theta^*+\cos^2\theta^*},
\end{equation}
       which at present is interpreted as a canonical signature for the
       intermediate vector (spin-1) bosons.
       The coefficient $A_{\rm FB}$ defines the backward-forward asymmetry,
       depending on $P$-parity of $Z'$ boson couplings to matter.

       In addition, another type of spin-1 bosons may exist, which leads to
       a different signature in the angular distribution.
       This follows from the presence of different types of relativistic spin-1 
       fermion currents $\bar{\psi}\gamma^\mu(1\pm\gamma^5)\psi$ and
       $\partial_\nu[\bar{\psi}\sigma^{\mu\nu}(1\pm\gamma^5)\psi]$, which
       can couple to the corresponding bosons.
       A clear example of such kind of interactions is provided by the
       hadron physics of the quark-antiquark mesons, which
       is considered as a low-energy QCD effective theory, where gluon and quark 
       degrees of freedom are substituted by physical hadronic states.
       It was pointed out
\cite{Chizhov:1996ky} that {\em three different} quantum numbers J$^{PC}$ 
      of existing neutral spin-1 mesons, ${1^{--}}$, ${1^{++}}$ and ${1^{+-}}$,
      cannot be assigned just to {\em two} vector $\bar{q}\gamma^\mu q$
      and axial-vector $\bar{q}\gamma^\mu \gamma^5 q$ quark states,
      which possess quantum numbers ${1^{--}}$ and ${1^{++}}$, respectively.
      The additional quark states
      $\partial_\nu(\bar{q}\sigma^{\mu\nu}q)$ and
      $\partial_\nu(\bar{q}\sigma^{\mu\nu}\gamma^5 q)$ are required,
      which also describe vector and axial-vector mesons,
      but with different transformation properties with respect to the Lorentz group
      and with different quantum numbers $1^{--}$ and $1^{+-}$, respectively.
      This example demonstrates that the pure tensor state, ${b_1}$ meson, exists.
      Furthermore, due to strong dynamics, vector $\rho$ and $\rho'$ mesons
      have minimal and anomalous couplings with vector
      $\bar{\psi}\gamma^\mu\psi$ and tensor
      $\partial_\nu(\bar{\psi}\sigma^{\mu\nu}\psi)$ interpolating currents
      comparable in magnitude 
\cite{Ball:1996tb,Braun:2003jg,Becirevic:2003pn,Chizhov:2003qy}.
      The both currents have the same quantum numbers J$^{PC}=1^{--}$ and mix.
      Since the parity and charge conjugation are
      conserved in QCD, they define the quantum numbers of the mesons.

      The mesons assigned to the tensor quark states are some types of
      ``excited'' states as far as the only orbital angular momentum with
      $L=1$ contributes to the total angular momentum, while the total spin
      of the system is zero.
      This property manifests itself in their derivative couplings 
      to matter and a different chiral structure of the anomalous 
      interactions in comparison with the minimal gauge ones.
      In contrast with the minimal gauge couplings,
      where either only left-handed or right-handed
      fermions participate in the interactions, the tensor currents mix
      both left-handed and right-handed fermions.
      Therefore, like the Higgs particles, the corresponding bosons
      carry a nonzero chiral charge.
      To our knowledge, such bosons were first introduced by Kemmer
\cite{Kemmer:1938zz} and they naturally appear in the extended conformal
      supergravity theories
\cite{Salam:1989fm}.
      In fact, this QCD feature can be realized in electroweak physics as well
      like the technicolor models.
      This analogy gives us arguments in favor of existence of
      anomalously interacting vector bosons.

      Up to now no search for excited bosons has been done, but regular
      searches for the excited lepton and quark states $f^*$
      have been carried out at the modern colliders, such as LEP
\cite{Barate:1998ce,Abbiendi:2002wf,Achard:2003hd}, HERA
\cite{Adloff:2002dy,Chekanov:2001xk} and Tevatron
\cite{Acosta:2004ri,Abazov:2008hw}.
      These excited fermions have magnetic moment (Pauli) type
      couplings to ordinary matter
\begin{equation}\label{f*}
    {\cal L}_{\rm excited}^{f^*}=
    \frac{g}{\Lambda}{\bar{f}^*}\sigma^{\mu\nu}\!f
    \left(\partial_\mu Z_\nu-\partial_\nu Z_\mu\right) +{\rm h.c.},
\end{equation}
      where the parameter $\Lambda$ is connected to the compositeness mass
      scale of the new physics.
      There are no objections to interpreting the interactions (\ref{f*})
      from a different point of view, introducing excited boson states
      instead of fermionic ones (``shifting the $*$-sign to the
      right from $f$ to $Z$'')
\begin{equation}\label{Z*}
    {\cal L}_{\rm excited}^{Z^*}=
    \frac{g}{\Lambda}\bar{f}\,\sigma^{\mu\nu}\!f
    \left(\partial_\mu{ Z^*_\nu}
    -\partial_\nu{ Z^*_\mu}\right).
\end{equation}
      This ``symmetry'' between excited fermions and bosons
      further supports our interest in consideration
      of the anomalously interacting vector bosons.
\smallskip

      This review paper summarizes our attempts
\cite{Chizhov:2008tp,Chizhov:2008jw,Chizhov:2010hh,Chizhov:2010jg,Chizhov:2011wt}
      to fill the gap in the consideration of experimental properties of
      heavy chiral bosons and continues discussions of possibilities 
      of disentangling them from other particles.

      The material below is given in the following order.
      In section II a simple chiral boson model is formulated for 
      further consideration.
      Section III contains our numerical estimations of the boson properties 
      within the framework of the CompHEP/CalcHEP package
\cite{Pukhov:1999gg,Boos:2004kh,Pukhov:2004ca}.
      To this end a new model has been implemented, which includes
      additional new bosons and their corresponding interactions.
      In section IV the first experimental constraints on the masses
      of both chiral vector bosons are given.
      Section V briefly describes new unique signatures
      of the bosons in dijet final states.
      Our conclusions are given in section VI.

\section{The chiral boson model}
      Let us assume that the electroweak gauge sector of the
      SM is extended by a doublet of new spin-1 {\em chiral\/}
      bosons $\mbox{\boldmath$W$}^*_\mu$ with the internal quantum numbers
      of the SM Higgs boson.
      There are at least three different classes of theories,
      all motivated by the Hierarchy problem, which predict
      new vector weak doublets with masses not far from the electro-weak scale.
      In particular, they can originate from the extensions of the
      SM such as Gauge-Higgs unification, larger gauge groups
      or technicolor models
\cite{Chizhov:2009fc}.
      However, due to the lack of fully realistic models, the collider
      expectations for signals from these chiral bosons have not yet been
      studied in detail.
      Nevertheless, it is possible to point out
      several model-independent and unique signatures which allow one
      to identify production of such bosons at the hadron colliders
\cite{Chizhov:2008tp}.

      Since the tensor current mixes the left-handed and right-handed
      fermions, which in the SM are assigned to different
      representations, the gauge doublet should have only anomalous
      interactions
\begin{equation}
    \label{master}
    {\cal L}^*=\frac{g}{M}\left(
                     \partial_\mu W^{*-}_\nu\;  \partial_\mu \overline{W}^{*0}_\nu
                 \right)\cdot
                 \overline{D_R}\;\sigma^{\mu\nu}\left(\hspace{-0.2cm}
                                                        \begin{array}{c}
                                                          U_L \\
                                                          D_L
                                                        \end{array}\hspace{-0.2cm}
                                                      \right)+
    \frac{g}{M}\left(
    \overline{U_L}\;  \overline{D_L}
                        \right)
                        \sigma^{\mu\nu}D_R\cdot
                        \left(\hspace{-0.2cm}
                          \begin{array}{c}
                            \partial_\mu W^{*+}_\nu \\
                            \partial_\mu W^{*0}_\nu \\
                          \end{array}\hspace{-0.2cm}\right),
\end{equation}
       where $M$ is the boson mass, $g$ is the coupling constant of the
       SU(2)$_{\rm W}$ weak gauge group, and $U$ and $D$ generically denote
       up-type and down-type leptons and quarks.
       This choice of couplings makes identical all partial fermionic 
       decay widths of the well-known hypothetical $W'$ boson
       with the SM-like interactions
\begin{equation}\label{W'}
{\cal
L}'_{CC}=\frac{g}{\sqrt{2}}\,W'^-_\mu\cdot\overline{D_L}\gamma^\mu
    U_L
    +\frac{g}{\sqrt{2}}\,\overline{U_L}\gamma^\mu D_L\cdot
    W'^+_\mu
\end{equation}
       and the charged $W^{*\pm}$ boson with the same mass. 
       Here we also assume universality of lepton and quark
       couplings with different flavors.
       In full analogy with the above-mentioned mesons
       these bosons, coupled to the tensor quark currents, can be considered
       as {\em excited\/} states.
       This property manifests itself in their derivative couplings
       to fermions and in the different chiral structure of the interactions
       in contrast to the minimal gauge interactions.

       For simplicity, in
(\ref{master}) we have introduced only interactions with the
       down-type right-handed singlets, $D_R$.
       In particular, in order to allow a possibility of
       detecting the neutral $CP$-even
       $Z^*=(W^{*0}+\overline{W}^{*0})/\sqrt{2}$ bosons
       via their decays into charged leptons
       (the Drell--Yan-like process)
       they should couple to the {\em down\/} type of fermions
\begin{equation}\label{Z*ed}{
    {\cal L}^{*}_{NC}=\frac{g}{2\sqrt{2} M}
    \left(\bar{\ell}\sigma^{\mu\nu}\ell+\bar{d}\sigma^{\mu\nu}d\right)
    \left(\partial_\mu Z^*_\nu-\partial_\nu Z^*_\mu\right)}.
\end{equation}
      Since we have introduced the complex
      $\mbox{\boldmath$W$}^*_\mu$ doublet, 
      there is an additional neutral $CP$-odd 
      $\tilde{Z}^*=(W^{*0}-\overline{W}^{*0})/\sqrt{2}$ boson. 
      However, in the case of light final states it is impossible 
      to discriminate the multiplicative quantum numbers of the 
      neutral bosons, namely $P$ and $C$.
      Therefore, in the following calculations
      we will consider only one of them, for instance, the $Z^{*}$ boson.


      For comparison we will consider topologically analogous gauge
      interactions of the $Z'$ boson
\begin{equation}\label{Z'ed}
{\cal L}'_{NC}=\frac{g}{2}
    \left(\bar{\ell}\gamma^{\mu}\ell+\bar{d}\gamma^{\mu}d\right)Z'_\mu
\end{equation}
      with the same mass $M$.
      The coupling constants are chosen in such a way that all fermionic 
      decay widths in the Born approximation of the both neutral
      bosons are identical.
      It means that their total production
      cross sections at the hadron colliders are nearly equal up to
      next-to-leading order corrections.
      Their total fermionic decay width
\begin{equation}\label{Gl}
    \Gamma=\frac{g^2}{4\pi}M\approx 0.034~M
\end{equation}
     is sufficiently narrow so that they can be identified as resonances
     at the hadron colliders in the Drell--Yan process.

    Furthermore, as several Higgs doublets are introduced 
    in many of the SM extensions, 
    the realistic model could include several gauge doublets.
    Using the charge-conjugated doublet
\begin{equation}\label{W*c}
    \mbox{\boldmath $W$}^{*\,{\rm c}}_\mu=\left(
             \begin{array}{c}
               \overline{W}^{*0}_\mu \\
               -W^{*-}_\mu \\
             \end{array}
           \right)
\end{equation}
        (or new ones with the hypercharges opposite to the $\mbox{\boldmath
        $W$}^*_\mu$ doublet) one can construct more complicated
         models including up-type right-handed singlets, $U_R$, as well.

\section{Numerical simulations of the chiral bosons}
    Up to now, any excess in the yield of the Drell--Yan process with
    high-energy invariant mass of the lepton pairs remains the clearest
    indication of possibile production of a new heavy neutral boson 
    at the hadron colliders.
    Therefore, we will first concentrate on consideration of
    the production and decay of neutral bosons,
    where full kinematics is experimentally reconstructible.
    In what follows we will use the CompHEP/CalcHEP package
\cite{Pukhov:1999gg,Boos:2004kh,Pukhov:2004ca} 
    for the numerical calculations of various
    distributions for the inclusive processes
$pp\rightarrow\gamma/Z/Z'+X\to\ell^+\ell^-+X$ and
$pp\rightarrow\gamma/Z/Z^*+X\to\ell^+\ell^-+X$
    with a CTEQ6M choice for the proton parton distribution set
    at $\sqrt{s}=10$~TeV.
    For both final leptons we impose angular restrictions (cuts) on the
    pseudorapidity range $|\eta_\ell|<2.5$ and the transverse momentum 
    $p_{\rm T} > 20$~GeV$/c$, which are relevant to the general LHC detectors.

\begin{figure}[htb]\center
\includegraphics[width=0.48\textwidth]{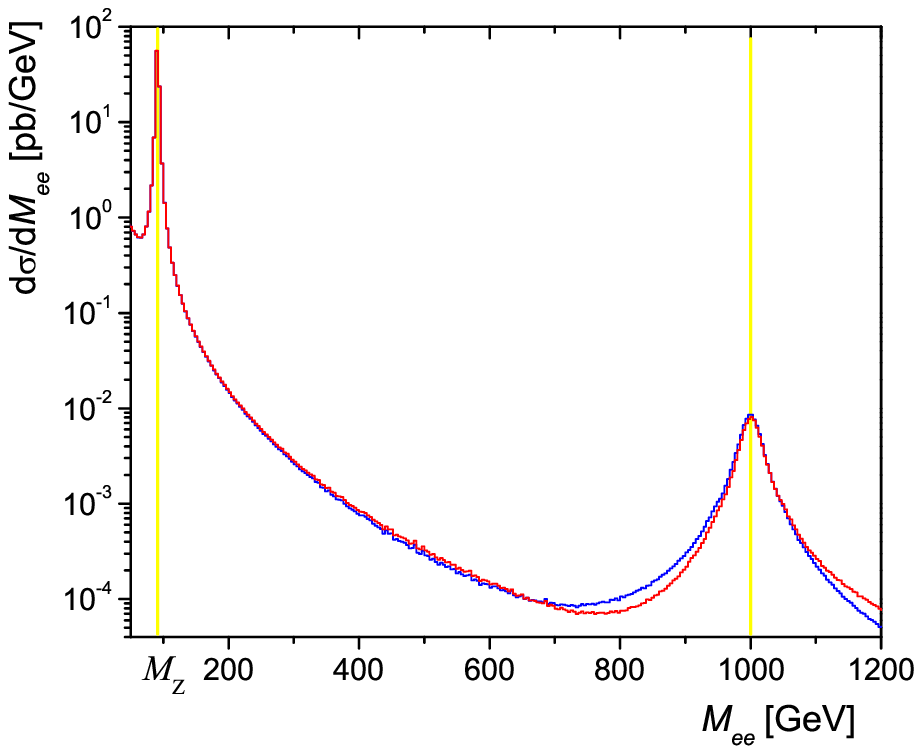}
\includegraphics[width=0.48\textwidth]{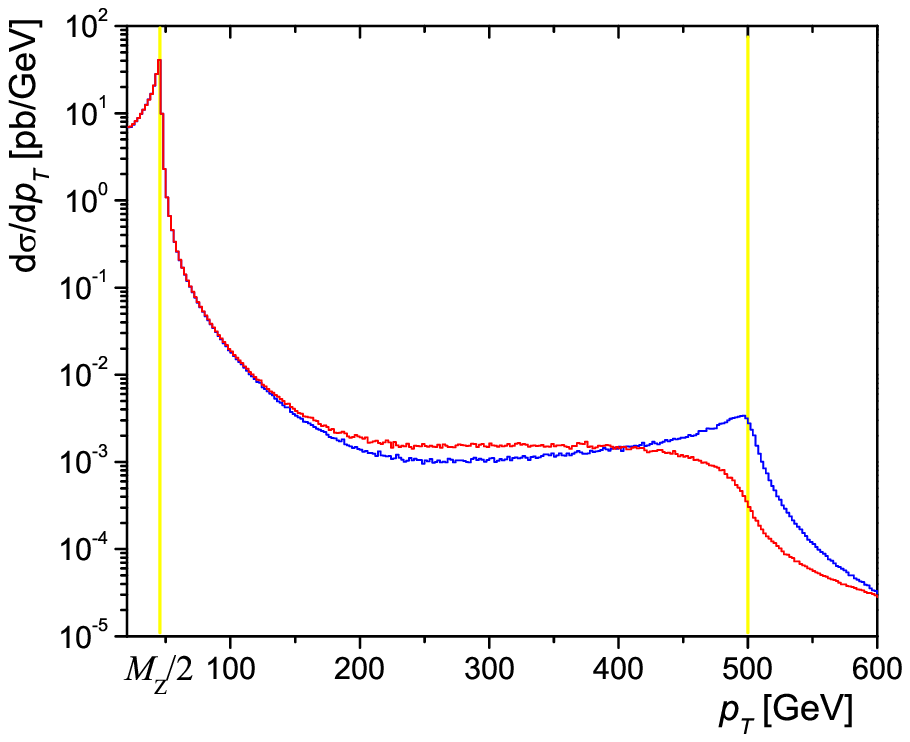}
\caption{ \label{fig:1-Mee} \label{fig:2}
    The invariant dilepton mass distributions (left) for the $Z'$ boson 
    (blue) and the excited chiral $Z^*$ boson 
    (red) with mass 1 TeV$/c^2$
    together with the Drell--Yan SM background (from the photon and the $Z$ boson)
    at the LHC for $\sqrt{s}=10$~TeV.
    The differential cross-sections (right) for the gauge $Z'$ boson (blue) and the
    excited chiral $Z^*$ boson (red) with the Drell--Yan SM background
    as functions of the lepton transverse momentum at the CERN LHC.}
\end{figure}

    Let us choose $M=1$~TeV/$c^2$ as a reference mass for new heavy
    bosons.
    For the high dilepton masses the cross sections of the new boson
    production with  
    this mass at the peak
    is about two orders of magnitude higher (in our model)
    than the corresponding Drell--Yan background, being the SM
    $\gamma$ and  $Z$ boson tails
    in the invariant dilepton mass    distributions
(Fig.~\ref{fig:1-Mee}, left).
       Therefore, the peak(s) should be clearly visible.

     The peaks in the dilepton invariant mass distributions originate from the
     Breit--Wigner propagator form, which is {\em the same}\/
     for both the gauge and chiral {\em neutral}\/
     bosons in the Born approximation.

     Concerning discovery of the {\em charged}\/
     heavy boson at the hadron colliders one believes that the cleanest method
     is detection of its subsequent leptonic
     decay into an isolated high transverse-momentum charged lepton
     (better without a prominent associated jet activity).
     In this case the heavy new boson can be observed
     through the Jacobian peak in the transverse 
     $p^{}_{\rm T}$ or $m^{}_{\rm T}$ distribution.
     It has become proverbial (see, for example, the textbook
\cite{Barger:1987nn}) that the Jacobian peak is an inevitable 
     characteristic of any two-body decay.
     However, it is not the case for decays of the new chiral bosons
\cite{Chizhov:2006nw}.
      It has been found in
\cite{Chizhov:2000vt} that tensor interactions lead to a new angular 
      distribution of the outgoing fermions
\begin{equation}
\label{GLR} 
    \frac{{\rm d} \sigma(q\bar{q}\to Z^*\!/W^*\to f\bar{f})}
     {{\rm d} \cos\theta} \propto \cos^2\theta,
\end{equation}
      in comparison with the well-known vector interaction result
\begin{equation}
\label{GLL}
    \frac{{\rm d} \sigma(q\bar{q}\to Z'\!/W'\to f\bar{f})}
    {{\rm d} \cos\theta} \propto     1+\cos^2\theta \, .
\end{equation}
      It was realized later
\cite{Chizhov:2006nw}
     that this property ensures a distinctive signature for the detection
     of the new interactions at the hadron colliders.
     At first sight, the small difference between the distributions
(\ref{GLR}) and
(\ref{GLL}) seems unimportant.
     However, the absence of the constant term in the first case results
     in very new experimental signatures.

     The angular distribution for vector interactions
(\ref{GLL})
     includes a nonzero constant term, which leads to the kinematical
     singularity in the $p^{}_{\rm T}$ distribution of the final fermion
\begin{equation}\label{1/cos}
    \frac{1}{\cos\theta}\propto\frac{1}{\sqrt{(M/2)^2-p^2_T}}
\end{equation}
    in the narrow width approximation $\Gamma <\!\!\!< M$
\begin{equation}\label{narrow}
    \frac{1}{(s-M^2)^2+M^2\Gamma^2}\approx\frac{\pi}{M\Gamma}\delta(s-M^2).
\end{equation}
     This singularity is transformed into a well-known Jacobian peak due
     to a finite width of the resonance.
     In contrast, the pole in
     the decay distribution of the $Z^*/W^*$ bosons is canceled out and
     the fermion 
     transverse momentum $p_{\rm T}$ distribution
     even reaches zero at the
     kinematical endpoint $p_{\rm T}=M/2$,
     rather than the Jacobian peak (at the kinematical
     endpoint $M/2$) for the gauge bosons
(Fig.~\ref{fig:2}, right).
     Therefore, even the lepton transverse momentum distribution
     demonstrates a difference between the gauge and chiral bosons.

\begin{figure}[htb]
\center
\vspace*{-0.75cm}\includegraphics[width=0.48\textwidth]{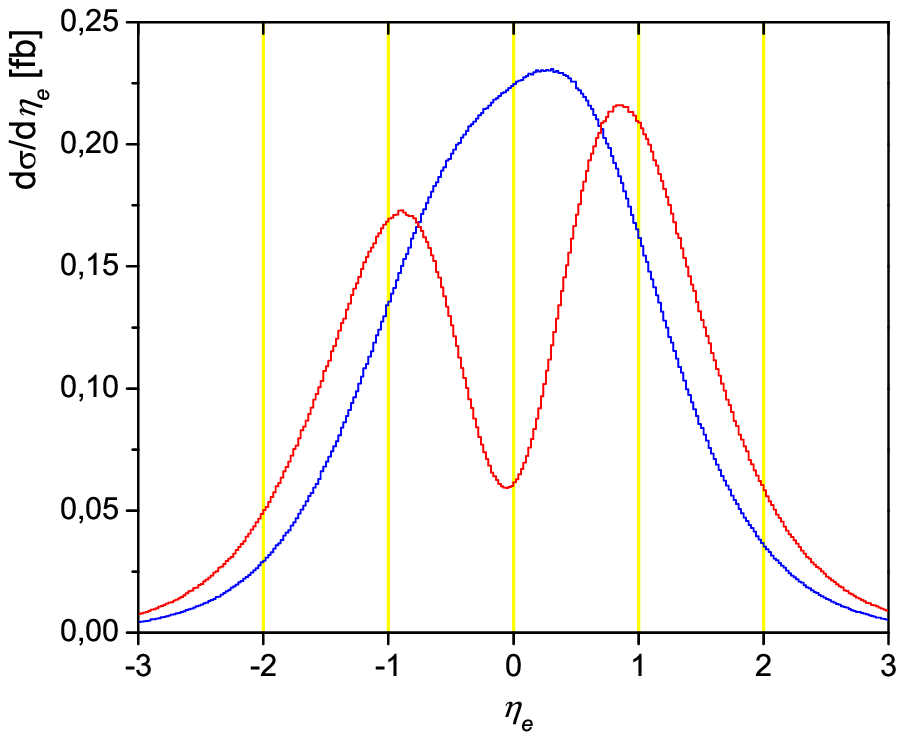}
\vspace*{-0.75cm}\includegraphics[width=0.48\textwidth]{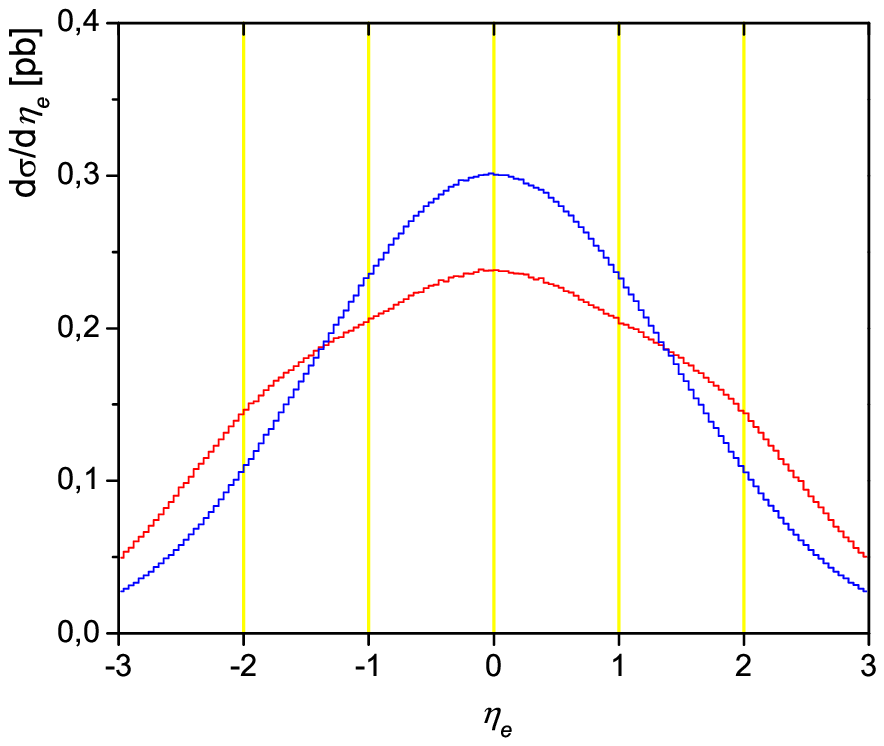}
\caption{
\label{fig:3}
\label{fig:4}
  The differential cross sections for the gauge $Z'$ boson (blue) and the
  excited chiral $Z^*$ boson (red) decaying to a lepton pair with the
  invariant mass 800~GeV/$c^2$~$<M_{\ell\ell}<$~1200~GeV/$c^2$ as functions of the
  lepton pseudorapidity at the Fermilab Tevatron 
  $\sqrt{s}=1.96$~TeV (left) and at the CERN LHC 
  $\sqrt{s}=14$~TeV
  (right).}
\end{figure}

      According to 
(\ref{GLR}), there exists a characteristic
      plane, perpendicular to the beam axis in the parton rest frame,
      where emission of final-state pairs is forbidden.
      The nonzero probability in the perpendicular direction in the laboratory
      frame is due to the longitudinal boosts of colliding partons.
      So, at the Fermilab Tevatron the production of such heavy bosons occurs
      almost at the threshold with approximately zero longitudinal momenta. 
      Hence, the lepton pseudorapidity distribution for the
      chiral bosons has a minimum at $\eta_\ell=0$
(Fig.~\ref{fig:3}, left).
      On the other hand, the CERN LHC is sufficiently powerful to produce
      heavy bosons with the mass $M=1$~TeV/$c^2$ with high longitudinal boosts.
      Therefore, the pseudorapidity distributions for the gauge and chiral
      bosons at the LHC look similar
(Fig.~\ref{fig:4}, right).

\begin{figure}[!h]
\vspace*{-0.3cm}
\centering
\includegraphics[width=0.48\textwidth]{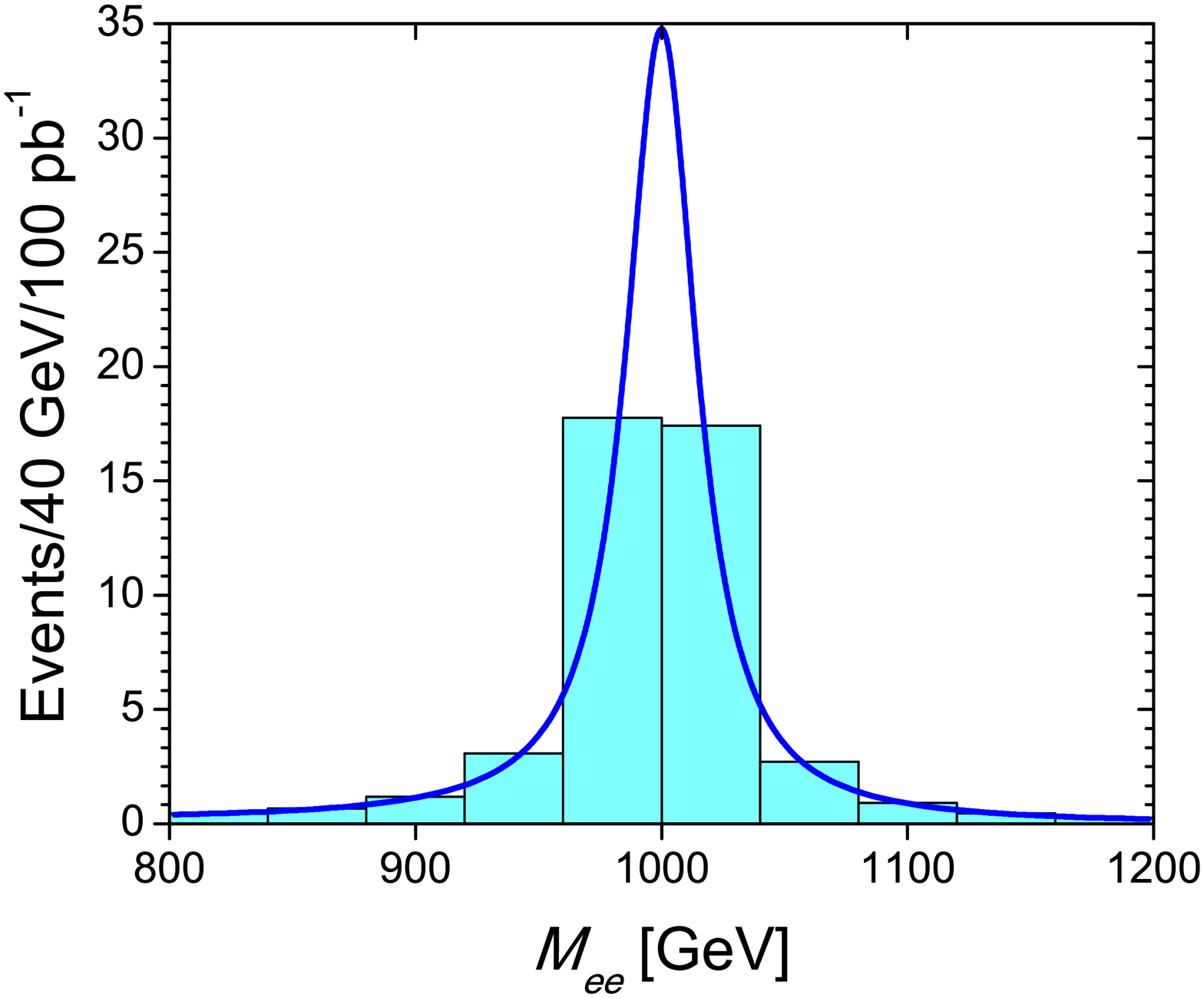}
\includegraphics[width=0.48\textwidth]{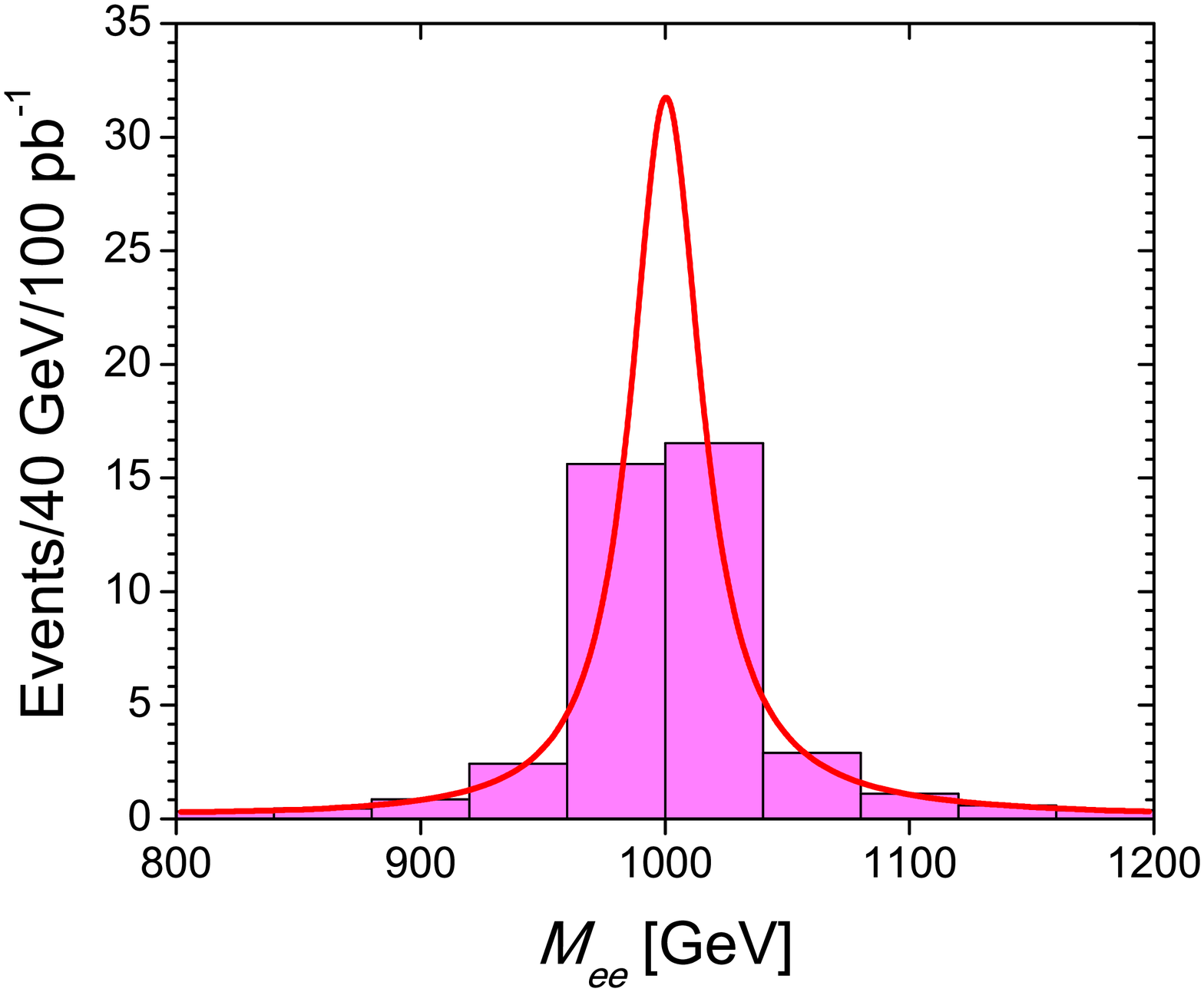}
\vspace*{-0.3cm}
\caption{
\label{fig:M_eeZZ}
     The invariant mass dilepton distributions for the gauge
     $Z'$ boson (left) and the excited chiral $Z^*$ boson (right)
     simulated with the common mass 1 TeV$/c^2$.
     They consist of 44.9 and 41.2 dilepton events in the mass window of
     800~GeV$/c^2<M_{\ell\ell}<$~1200~GeV$/c^2$
     for $Z'$ 
     and $Z^*$ bosons.  
}
\end{figure}

     In order to make more substantial and experiment-looking conclusions,
     let us investigate signal 
     distributions selecting only ``on-peak'' events with the invariant
     dilepton masses in the range
800~GeV$/c^2<M_{\ell\ell}<$~1200~GeV$/c^2$.
     To this end (as a test example) in
\cite{Chizhov:2010hh}
     for these dilepton masses and kinematical restrictions
     $|\eta_\ell|<2.5$ and the transverse lepton momentum
     $p_{\rm T} > 20$~GeV$/c$
     we have simulated production of dilepton events via
$pp\rightarrow\gamma/Z/Z'+X\to\ell^+\ell^-+X$ and
$pp\rightarrow\gamma/Z/Z^*+X\to\ell^+\ell^-+X$
      for the LHC integrated luminosity of 100~pb$^{-1}$
      and $\sqrt{s}=10$~TeV.
      As far as the center-of-mass energy for the 2010--2011 runs was
      7~TeV, at which the cross sections are roughly half as large, 
      200~pb$^{-1}$ of data will be equivalent to the case with
      $\sqrt{s}=10$~TeV.
      For the $Z'$ and $Z^*$ bosons (with mass 1 TeV$/c^2$)
      production cross-sections
      $\sigma_{Z'}=0.45$~pb and
      $\sigma_{Z^*}=0.41$~pb were obtained.
      The values respectively transform 
      into 44.9 and 41.2 dilepton events in the mass window of
      800~GeV$/c^2<M_{\ell\ell}<$~1200~GeV$/c^2$
(see Fig.~\ref{fig:M_eeZZ}).
      Under the same kinematical conditions the SM gives
      0.6 events with the production cross-section 5.75 fb.

\begin{figure}[htb]\center
\includegraphics[width=0.48\textwidth]{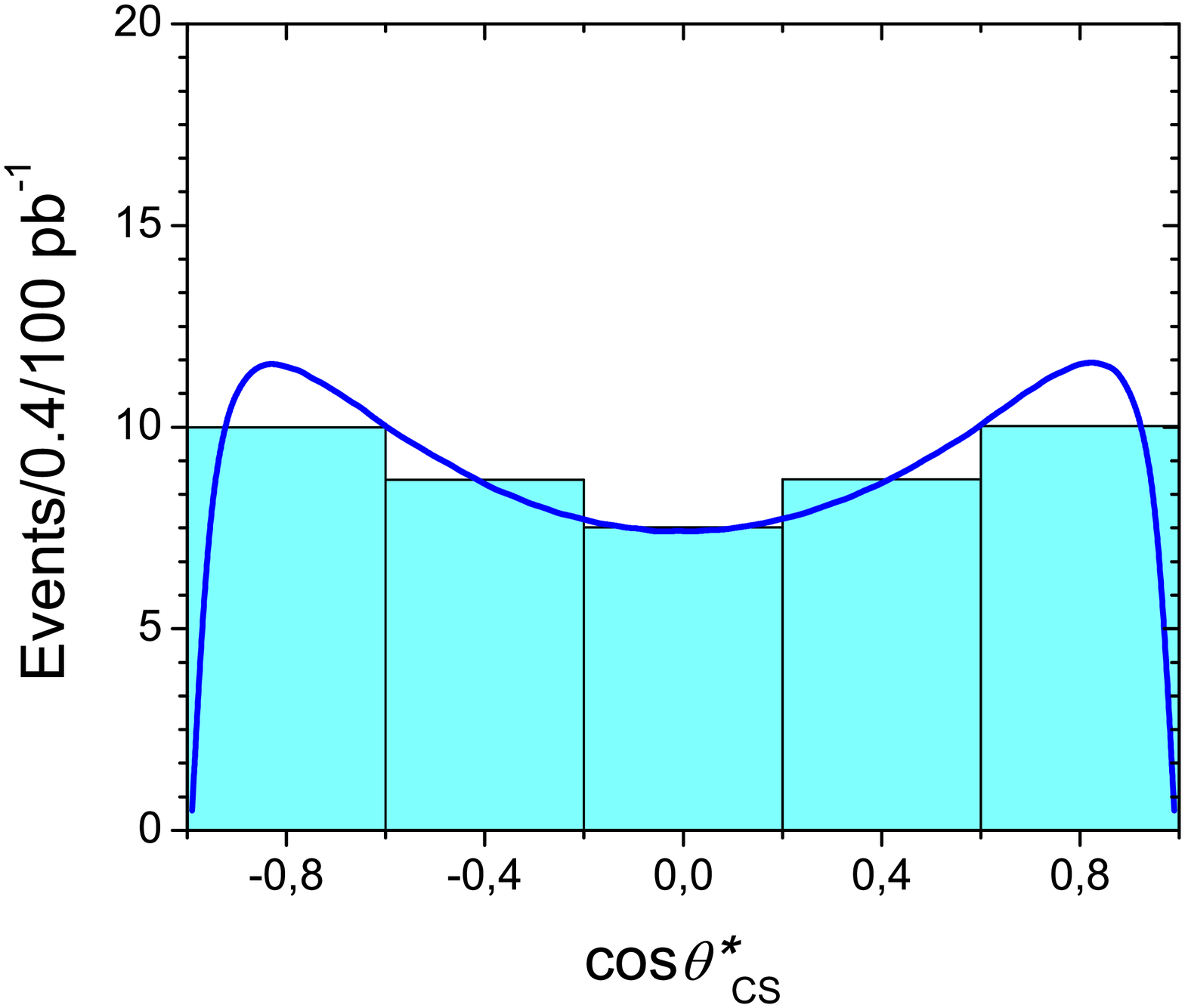}
\includegraphics[width=0.48\textwidth]{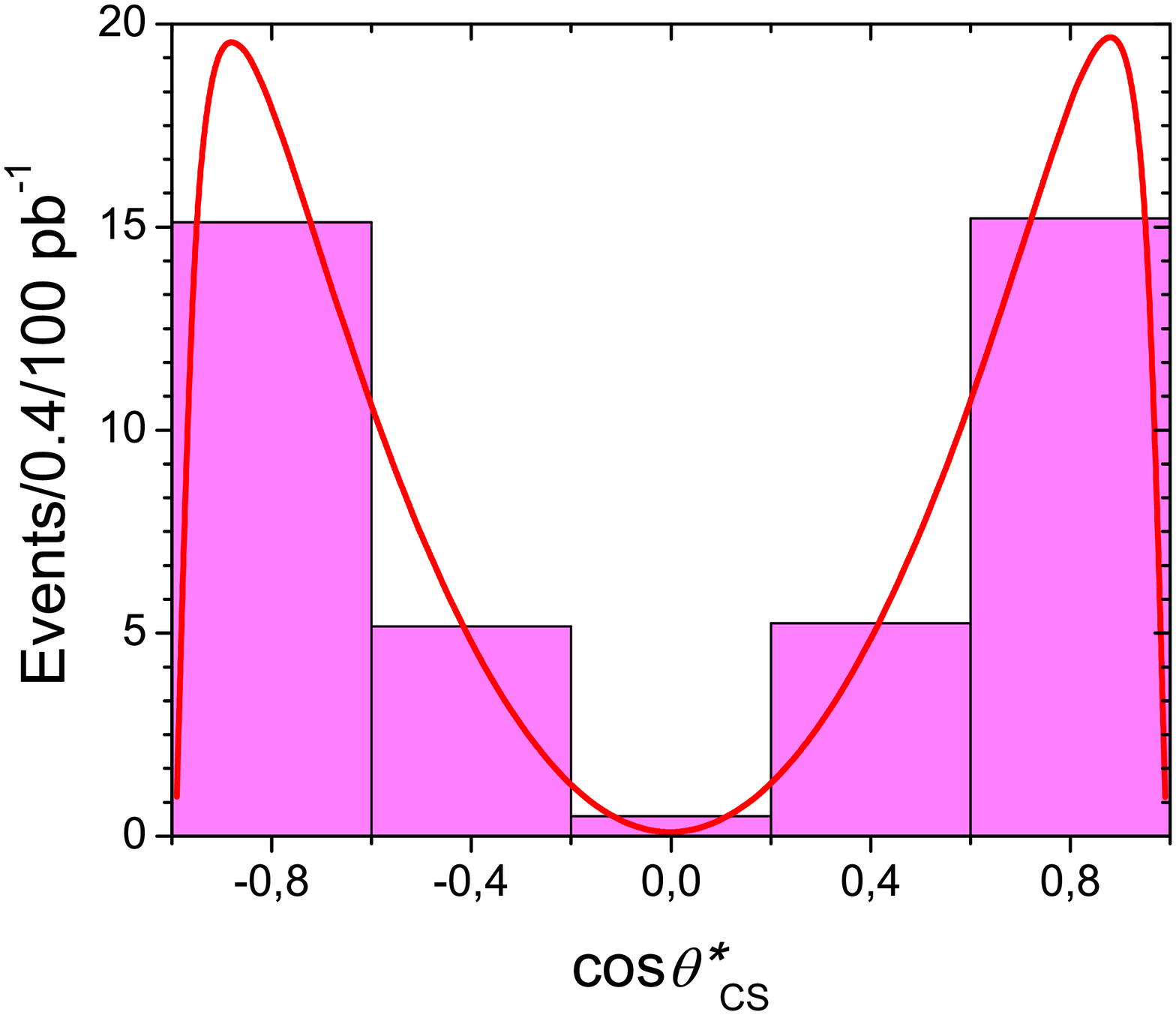}
\caption{\label{fig:CS}
       The differential lepton angular distributions of the gauge
       $Z'$ boson (left) and the excited chiral $Z^*$ boson (right) as
       functions of $\cos\theta^*_{\rm CS}$ for $M=1$~TeV/$c^2$.
}
\end{figure}

       As already mentioned,
       the peaks in the invariant mass distributions
       originate from the Breit--Wigner propagator form, which is the same
       for both $Z'$ and $Z^*$ bosons in the leading Born approximation.
       Therefore, in order to discriminate them we need to investigate
       additional distributions selecting only ``on-peak'' events with the
       invariant dilepton masses in the 
       optimal window size $[M-2\Gamma,M+2\Gamma]$. 
       According to 
\cite{Chizhov:2008tp}, a crucial difference between the 
        neutral chiral bosons
        and other resonances should come from the analysis of the angular
	distribution of the final-state leptons with respect to the boost
	direction of the heavy boson in the rest frame of the latter (the
	Collins--Soper frame
\cite{Collins:1977iv})
(Fig.~\ref{fig:CS}).
       Instead of a smoother angular distribution for the gauge
       interactions (left),
       a peculiar ``swallowtail'' shape of the chiral boson
       distribution (right) occurs with a dip at $\cos\theta^*_{\rm CS}=0$.
       It will indicate the presence of the new interactions.
       Neither scalars nor
       other particles possess such a type of angular behavior.
       Indeed, the angular distribution of outgoing leptons
       for the $Z^*$ bosons will lead to the large negative value
       of the so-called centre-edge asymmetry $A_{\rm CE}$:
\begin{equation}\label{ACE}
    \sigma\times A_{\rm CE}=\hspace{-0.2cm}\int^{+\frac{1}{2}}_{-\frac{1}{2}}
    \hspace{-0.2cm}\frac{\drm\sigma}{\drm\cos\theta^*_{\rm CS}}\,\drm\cos\theta^*_{\rm CS}-
    \hspace{-0.1cm}\left[\int^{+1}_{+\frac{1}{2}}
    \hspace{-0.2cm}\frac{\drm\sigma}{\drm\cos\theta^*_{\rm CS}}\,\drm\cos\theta^*_{\rm CS}+
    \hspace{-0.2cm}\int^{-\frac{1}{2}}_{-1}
    \hspace{-0.3cm}\frac{\drm\sigma}{\drm\cos\theta^*_{\rm CS}}\,\drm\cos\theta^*_{\rm CS}
    \right]\!,
\end{equation}
\begin{figure}[htb]\center
\includegraphics[width=0.48\textwidth]{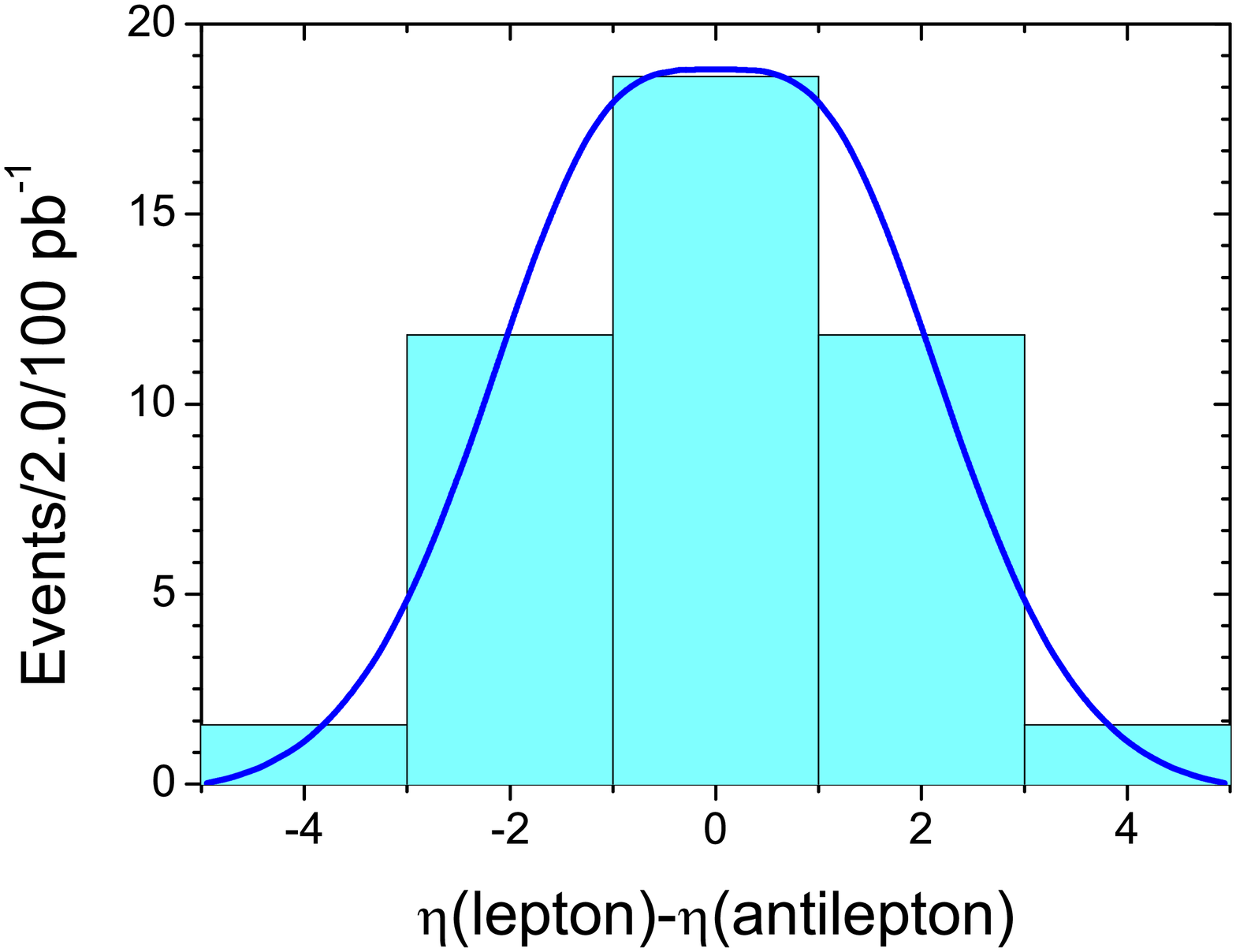}
\includegraphics[width=0.48\textwidth]{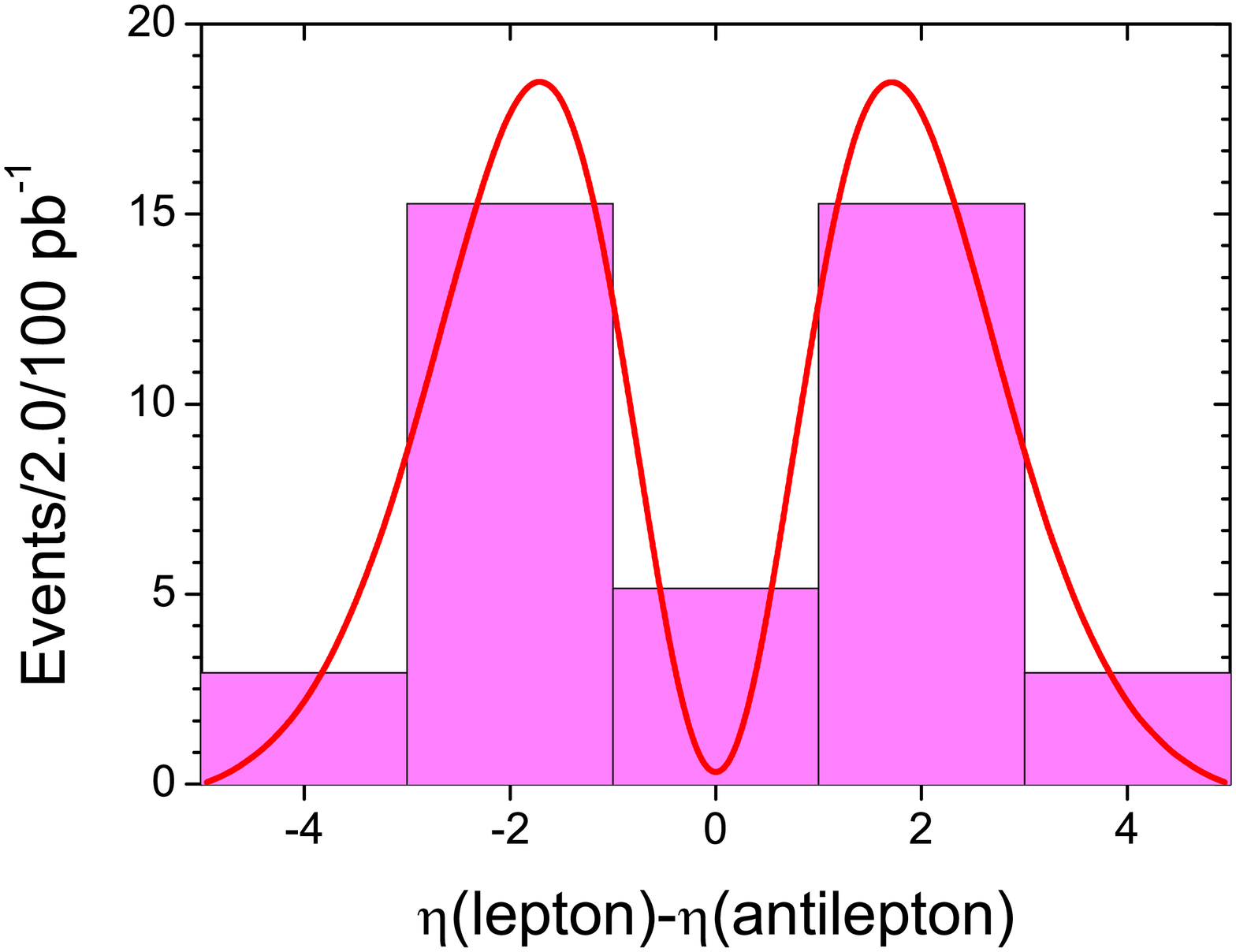}
\caption{\label{fig:eta}
      The differential distributions for the
      gauge $Z'$ boson (left) and the excited chiral $Z^*$ boson (right)
      as functions of the difference of the lepton pseudorapidities for $M=1$~TeV/$c^2$.}
\end{figure}
       while the distributions of other known resonances (even with
       different spins) possess positive or near-zero asymmetries.
%
       Using this asymmetry, one can
       strongly reduce the systematic uncertainties from the hadron structure
\cite{Gounaris:1992kp}.

       Another ``unexpected'' consequence of the new
       form of angular distribution
(\ref{GLR}) is a very different shape of the event distribution over
       pseudorapidity difference 
($\eta_1-\eta_2$) between both outgoing charged leptons.
       It is shown in
Fig.~\ref{fig:eta}.
       Combining these distributions one will have a possibility 
       of differentiating these bosons for higher resonance masses.

\begin{figure}[htb]\center
\includegraphics[width=0.48\textwidth]{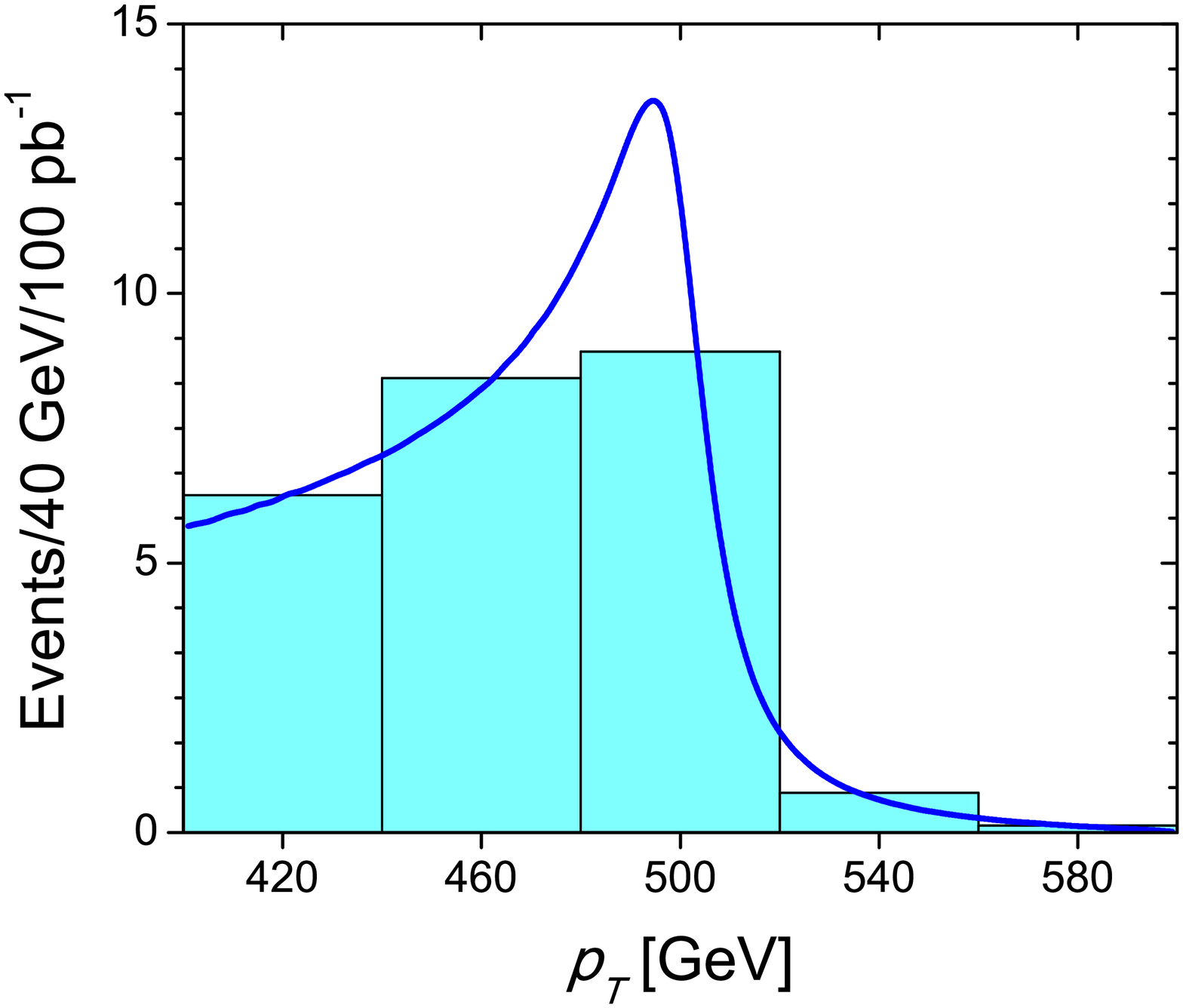}
\includegraphics[width=0.48\textwidth]{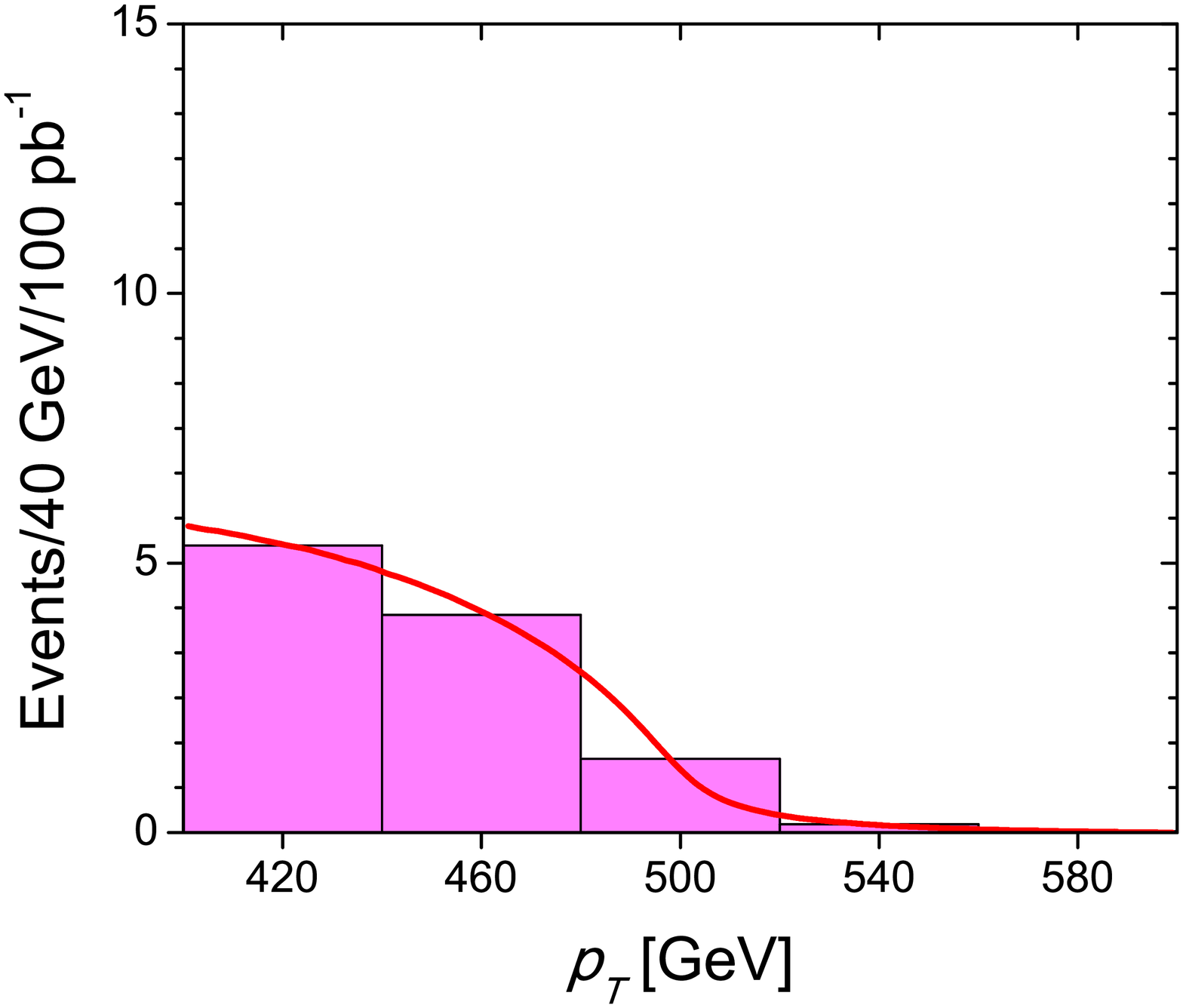}
\caption{\label{fig:pT}
       The differential distributions for the $Z'$
       boson (left) and the chiral excited $Z^*$ boson (right) as functions
       of the lepton transverse momentum $p_{\rm T}$ for
       $M=1$~TeV/$c^2$
       the LHC integrated luminosity of
       100~pb$^{-1}$ and $\sqrt{s}=10$~TeV.}
\end{figure}

       We would like again to consider the 
       distributions of the lepton transverse momentum $p_{\rm T}$.
       As mentioned before and demonstrated for our simulated sample in
Fig.~\ref{fig:pT},
     the relevant $Z^*$ and $W^*$ boson decay distributions have
     a broad smooth hump with the maximum below the kinematical endpoint,
     instead of an ``expected'' sharp Jacobian peak.
     Therefore, in contrast to the usual procedure
     of the direct and precise determination of the 
     $W'$ resonance mass, the new distribution does not allow doing 
     it for $W^*$ bosons.
     Moreover, even a relatively small decay width of the chiral bosons
     has a wide $p_{\rm T}$ distribution
     that obscures their identification as resonances at hadron
     colliders.

\section{The first experimental constraints on the chiral bosons}

\def\zpssm{\ensuremath{Z'_{\rm SSM}}}
\def\zppsi{\ensuremath{Z'_\psi}}
\def\zpchi{\ensuremath{Z'_\chi}}
\def\zpeta{\ensuremath{Z'_\eta}}
\def\zpsq{\ensuremath{Z'_{\rm S}}}
\def\zpN{\ensuremath{Z'_{\rm N}}}
\def\zpI{\ensuremath{Z'_I}}
\def\zpiota{\ensuremath{Z'_\iota}}
\def\ll{{\ensuremath{\ell^+\ell^-}}}
\def\ee{{\ensuremath{e^+e^-}}}
\def\mumu{{\ensuremath{\mu^+\mu^-}}}

\def\zstar{\ensuremath{Z^*}}
\def\zpll{\ensuremath{Z'\rightarrow \ell ^+ \ell ^-}}
\def\zsll{\ensuremath{Z^*\rightarrow \ell ^+ \ell ^-}}
\def\zpee{\ensuremath{Z'\rightarrow e^+ e^-}}
\def\zsee{\ensuremath{Z^*\rightarrow e^+ e^-}}
\def\zpmumu{\ensuremath{Z'\rightarrow \mu ^+ \mu ^-}}
\def\zsmumu{\ensuremath{Z^*\rightarrow \mu ^+ \mu ^-}}

\def\met{\ensuremath{E^{\rm miss}_{\mathrm T }}}

\def\mt{\ensuremath{m_{\mathrm T}}}
\def\et{\ensuremath{E_{\mathrm T}}}
\def\pt{\ensuremath{p_{\mathrm T}}}
\def\wp{\ensuremath{W'}}
\def\wstar{\ensuremath{W^*}}
\def\mettext{missing \et}
\def\tauleps{$\tau$-leptons}
\def\xbr{\ensuremath{\sigma B}}
\def\wps{\ensuremath{W'/W^*}}
\def\wpse{\ensuremath{\wps\rightarrow e \nu}}
\def\wpsmu{\ensuremath{\wps\rightarrow \mu \nu}}

\begin{figure}[!htb]
\hspace*{-0.6cm}
\includegraphics[width=0.5\columnwidth]{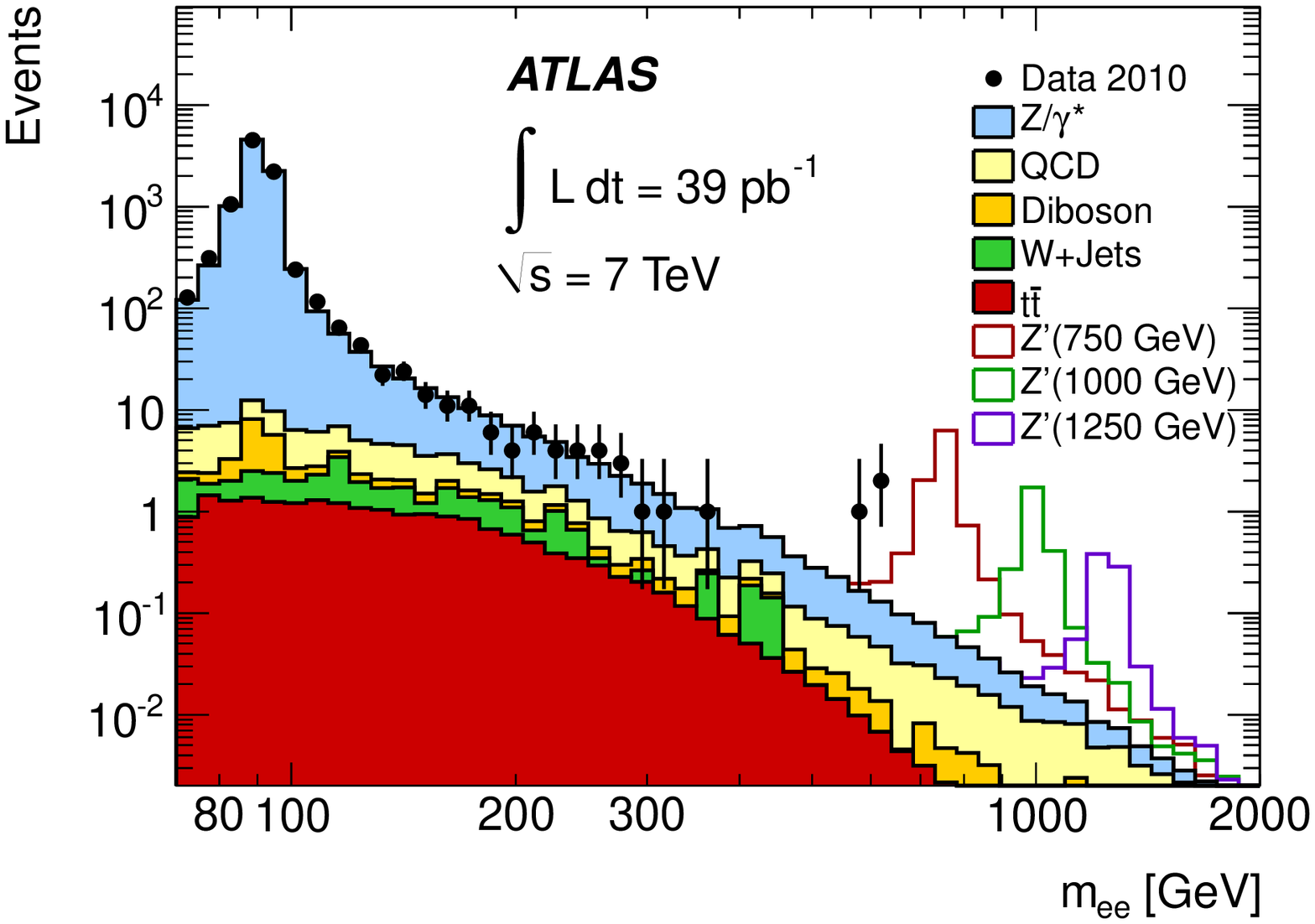}
\includegraphics[width=0.5\columnwidth]{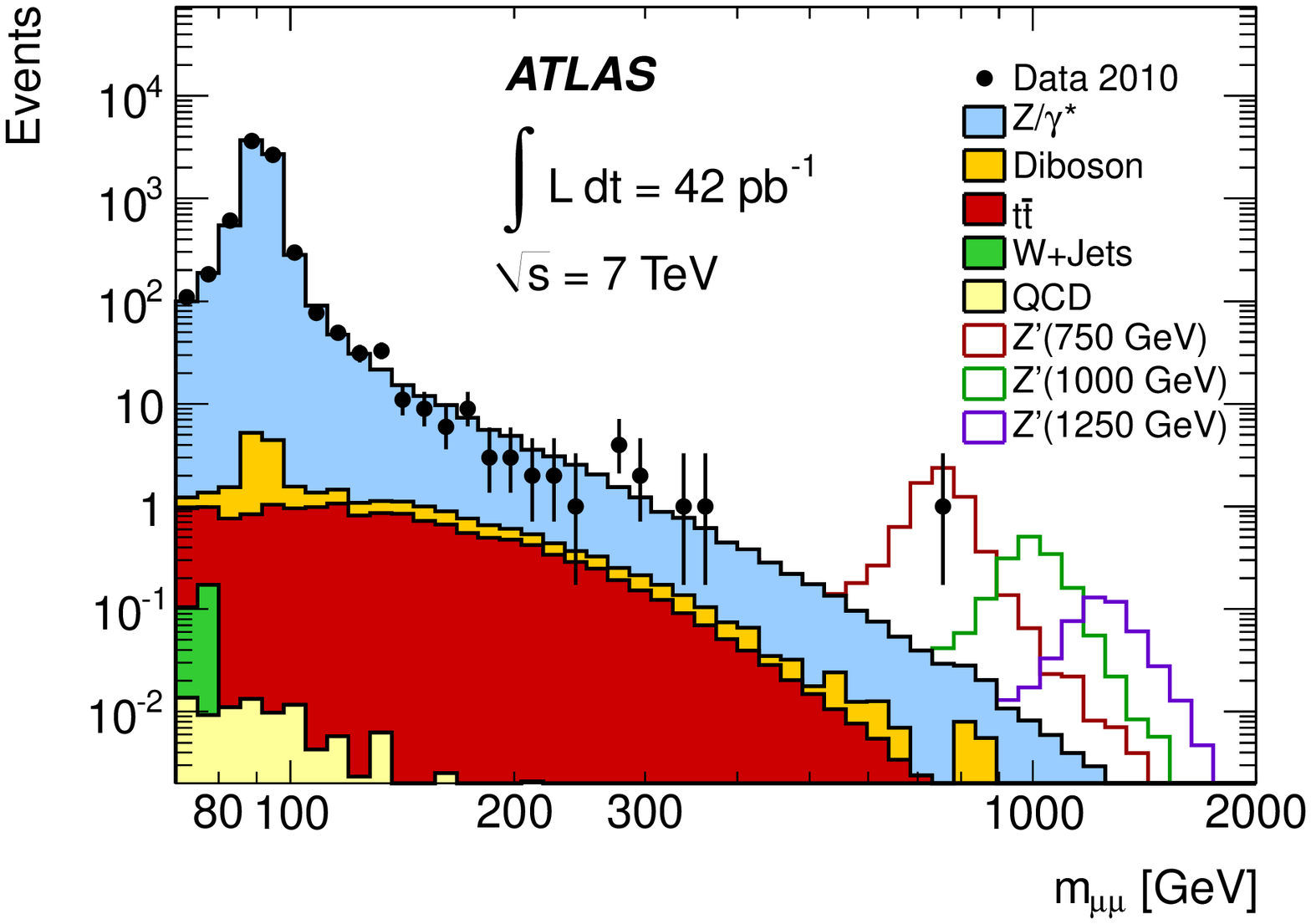}
  \caption{Dielectron (left) and dimuon (right)
          invariant mass distribution measured by the ATLAS collaboration in 2010
\cite{Aad:2011xp}.
          They are compared with all expected backgrounds
      and three example $Z'$ signals.
}  \label{fig:init_res3}
   \label{fig:moneyPlot}
\end{figure}
\begin{figure}[!htb]
\hspace*{-0.6cm}
\includegraphics[width=0.5\columnwidth]{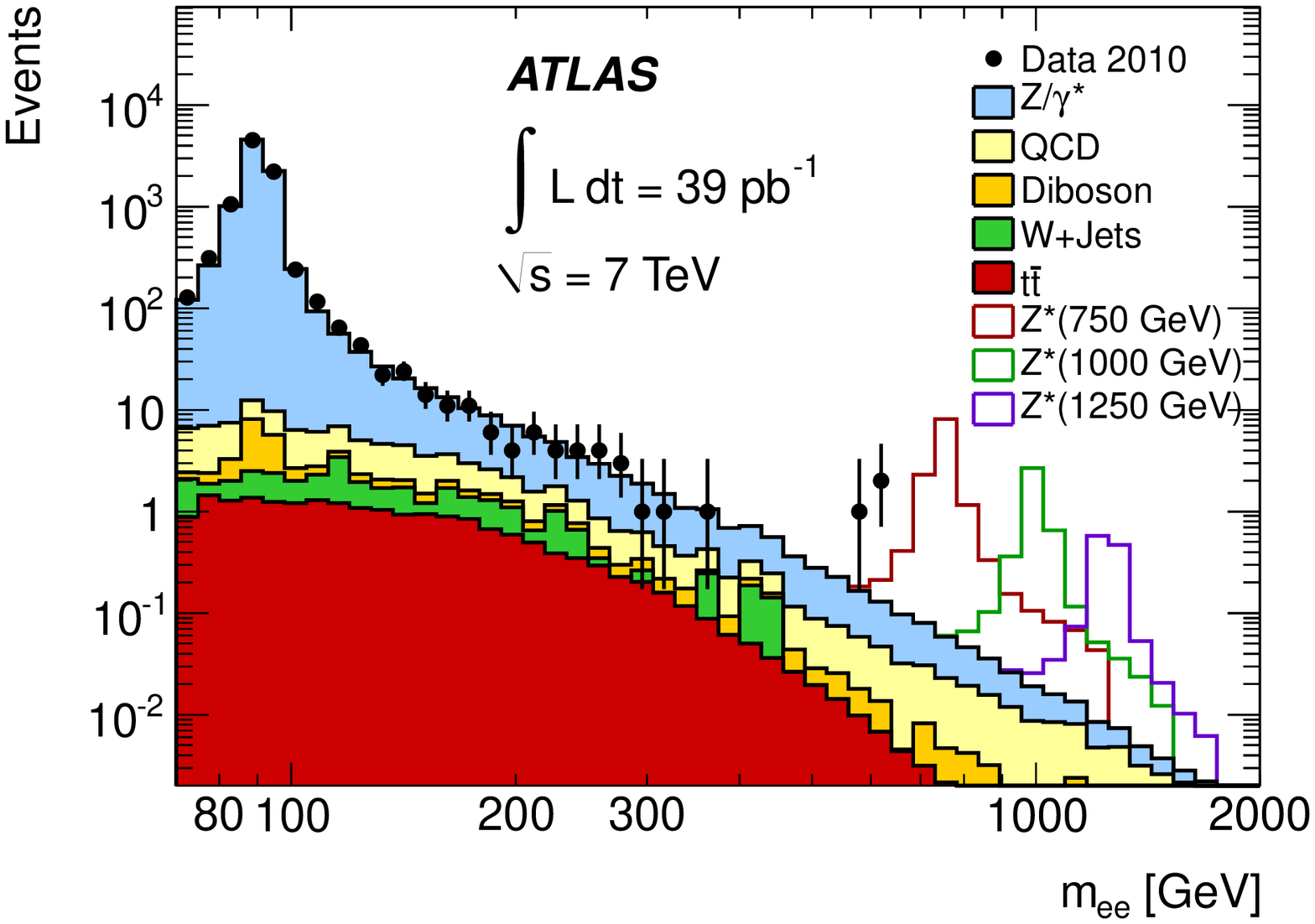}
\includegraphics[width=0.5\columnwidth]{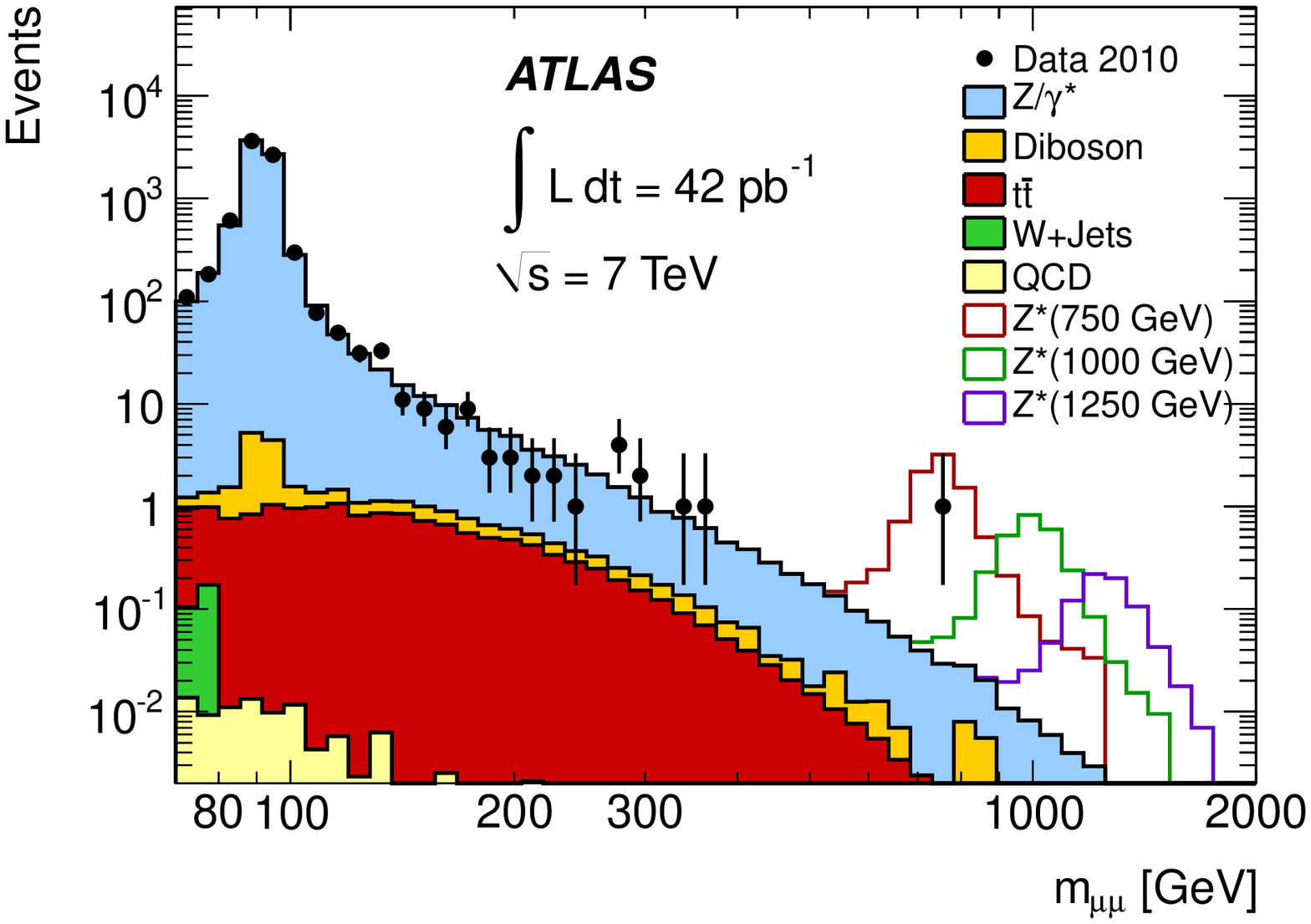}
  \caption{Dielectron (left) and dimuon (right)
          invariant mass distribution after final selection,
          compared to the stacked sum of all expected backgrounds,
          with three example $Z^*$ signals overlaid.
}
   \label{fig:ZstarPlots}
\end{figure}

      The first direct experimental search for the excited chiral vector bosons
      was performed 
      by the ATLAS collaboration
\cite{Aad:2008zzm,Collaboration:2010knc,Aad:2011dr} in 2010.
      At the LHC energy of 7 TeV with the 
      integral luminosity around 40~pb$^{-1}$
      the ATLAS detector was used for searching for narrow resonances
      in the invariant mass spectrum above 110~GeV$/c^2$
      of $e^+e^-$ and $\mu^+\mu^-$ final states.
      The main physical results of the relevant paper
      ``Search for high mass dilepton resonances in
      $pp$ collisions at $\sqrt{s}=7$~TeV with the ATLAS experiment''
\cite{Aad:2011xp} are presented in 
Fig.~\ref{fig:init_res3}
      together with main backgrounds and
      expected $Z'$ decay signals for three masses around 1~TeV$/c^2$.
      Expected signals from the  $Z^*$ boson decays considered in this paper
      (are shown in Fig.~\ref{fig:ZstarPlots})
      have similar shapes and approximately 40\% larger cross sections.
      Three ``interesting'' events in the vicinity of $m_{ee} = 600$~GeV$/c^2$
      and a single event at $m_{\mu\mu} = 768$~GeV$/c^2$ are observed.
      All details of the data selection and the physical data analysis
      can be found in
\cite{Aad:2011xp}.
      It is seen that both the dielectron and dimuon invariant 
      mass distributions are well described by the prediction 
      from SM processes.
      From these figures one can conclude that no statistically 
      significant excess above the SM  expectation is observed 
      with these data samples.
      Nevertheless, these distributions were 
      for the first time used to
      obtaine a lower direct mass limit (1.152~TeV$/c^2$, see below)
      for the neutral chiral $Z^*$ boson described in this paper.
      To this end, 95\% C.L. exclusion limits on 
      $Z^*$ production and its dilepton decay, ${\sigma B}$,
      for the combination of the electron and muon
      $Z^*$ boson decay channels were used.
       The combination was performed by defining the likelihood
       function in terms of the total number of
       $Z^*$  events produced in both channels.

       In the three cases (dielectron, dimuon and combined channels),
       the 95\% C.L. ${\sigma B}$ limit 
       was used to set mass limits for each of the considered models.
       The observed combined mass limit for the Sequential Standard Model
       \zpssm\ is 1.048~TeV$/c^2$. 
       The limits on the E6-motivated $Z'$ bosons
       are in the range 0.738--0.900 TeV$/c^2$.
       Although the angular lepton decay distributions are not the same
       for $Z'$ and $Z^*$ bosons, it was found that
       the difference in geometrical acceptance is negligible
       for the boson pole masses above 750~GeV$/c^2$.
       Therefore, the same procedure as for the $Z'$ bosons
       is used to calculate the limit on $\xbr (\zsll )$
       and on the $Z^*$ boson mass in each channel and for their combination.
       Finally the observed combined
       lower mass limit for the $Z^*$ boson is 1.152~TeV$/c^2$.
       This is the first direct mass limit on this particle.
       The $Z^*$ limits are about 100--200 GeV/$c^2$ more stringent than
       the corresponding limits on all considered $Z'$ bosons.
\smallskip

\begin{figure*}[!h]
  \centering
  \includegraphics[width=0.50\textwidth]{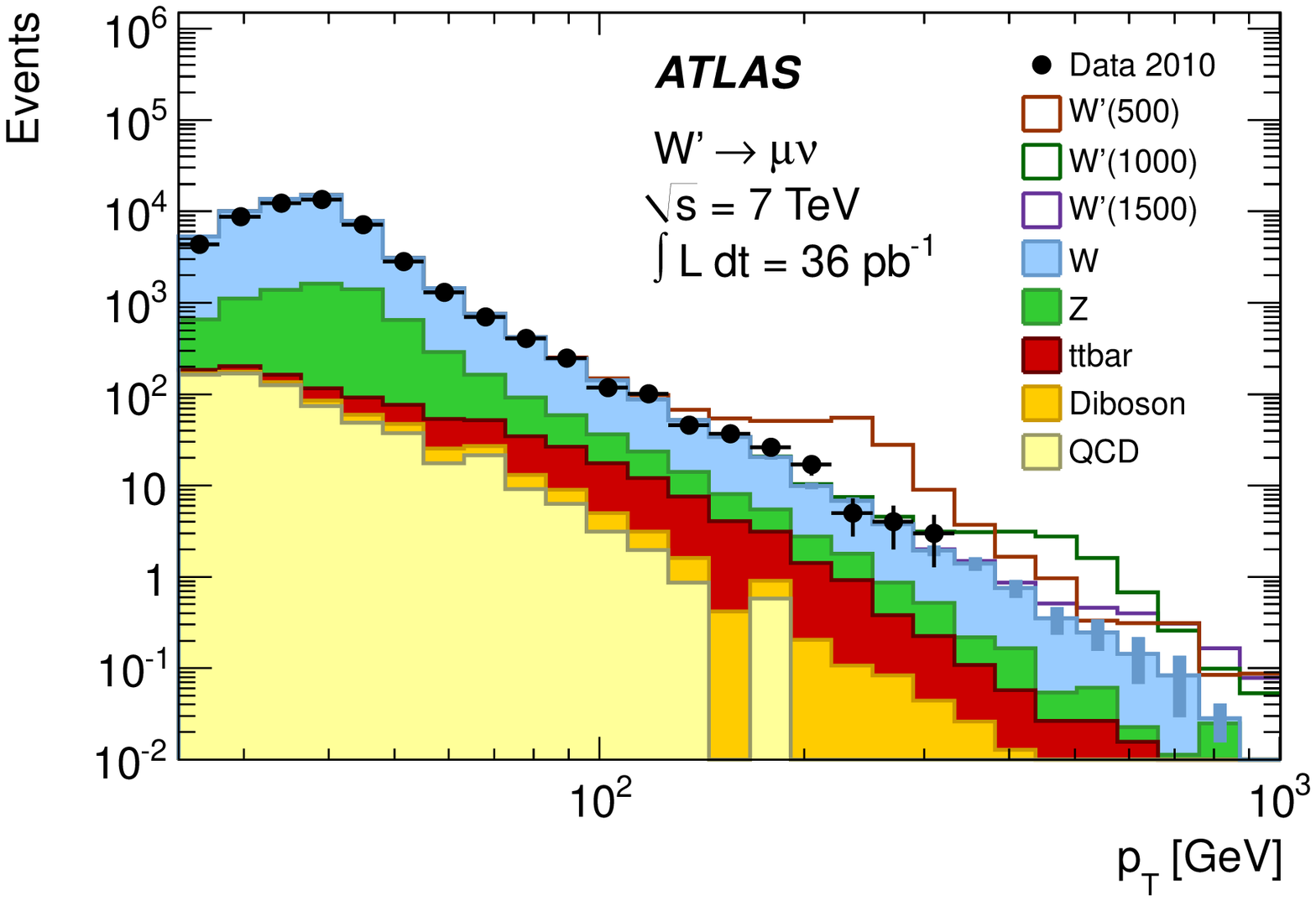}
  \includegraphics[width=0.43\textwidth]{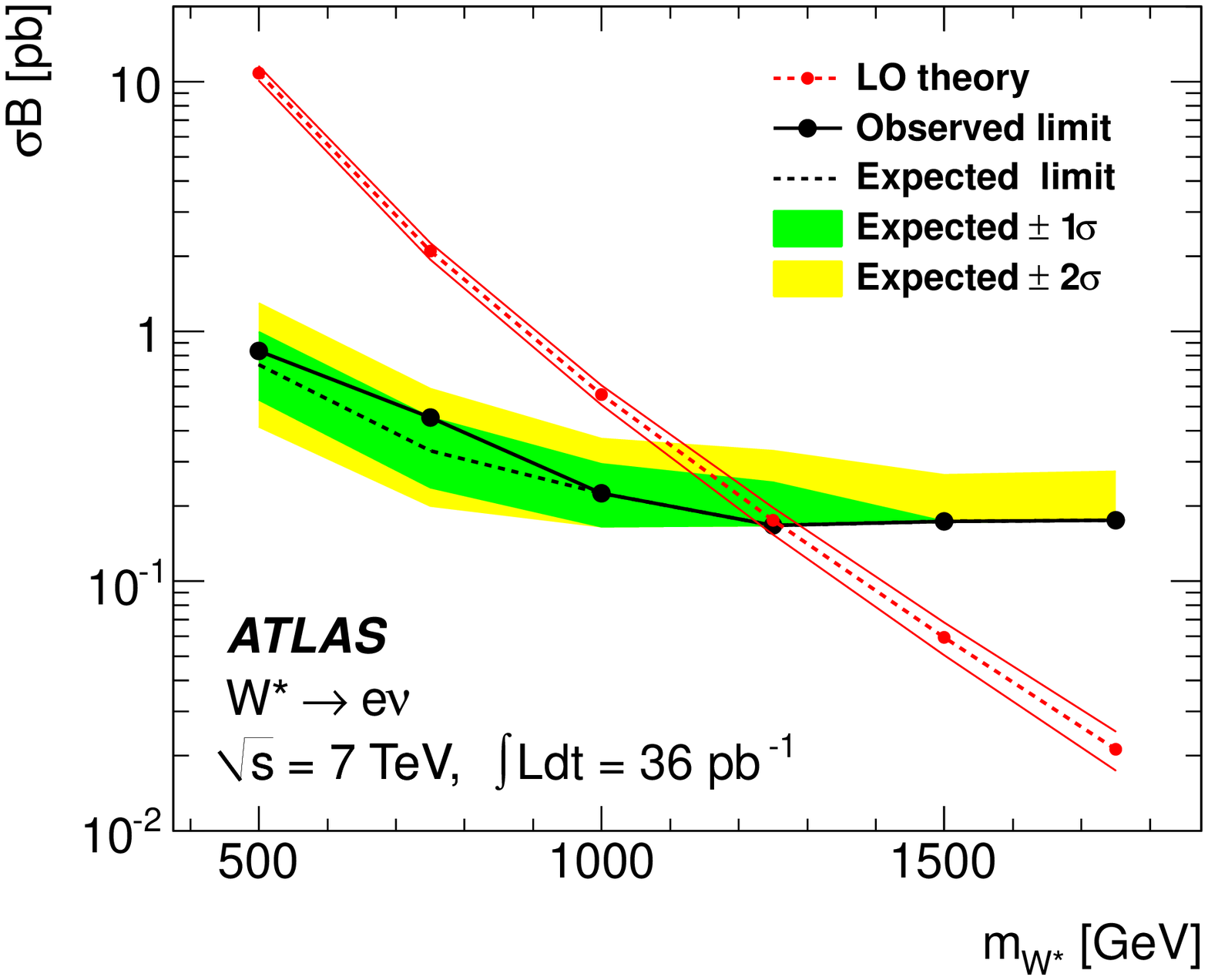}
  \includegraphics[width=0.50\textwidth]{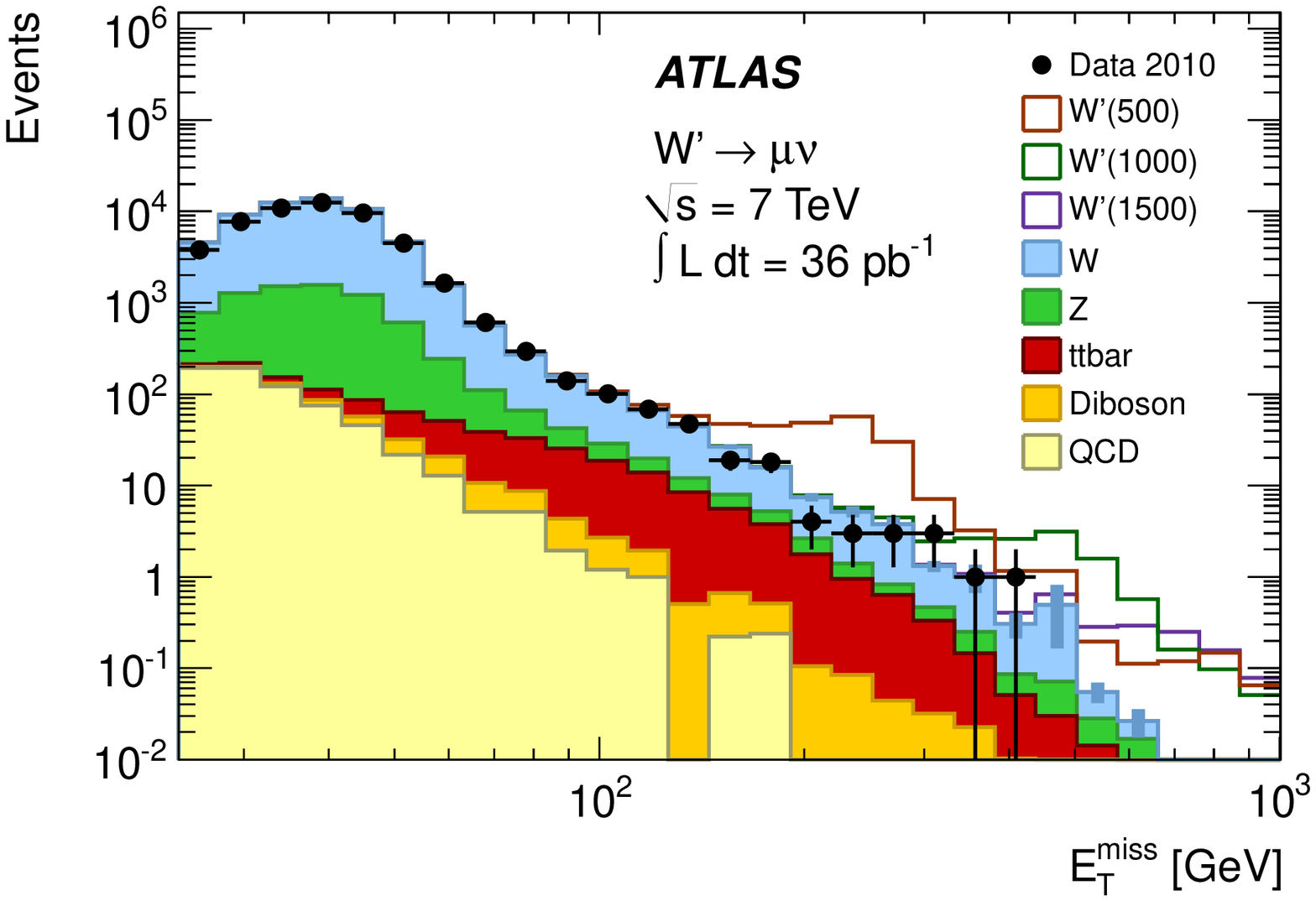}
  \includegraphics[width=0.43\textwidth]{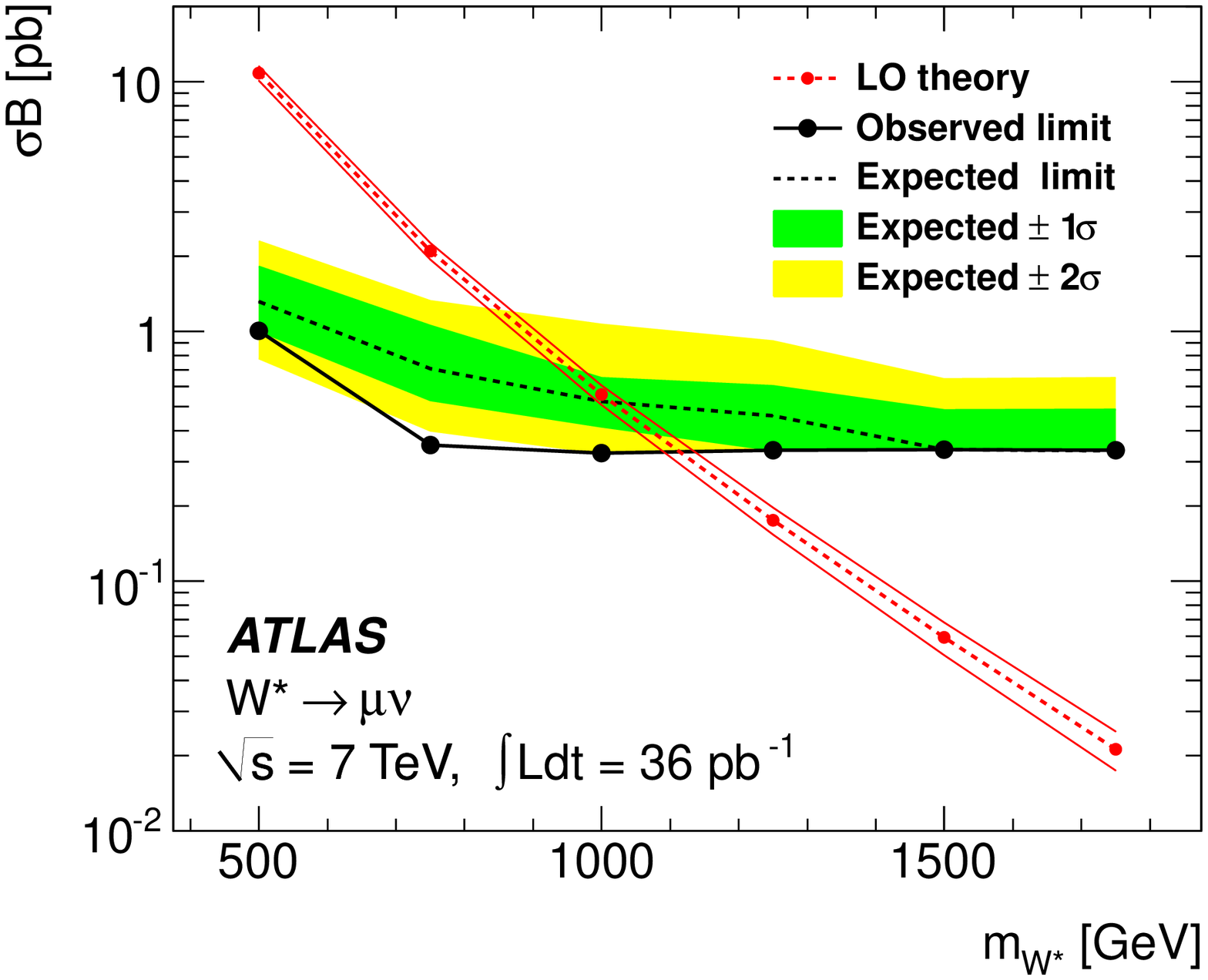}
  \includegraphics[width=0.50\textwidth]{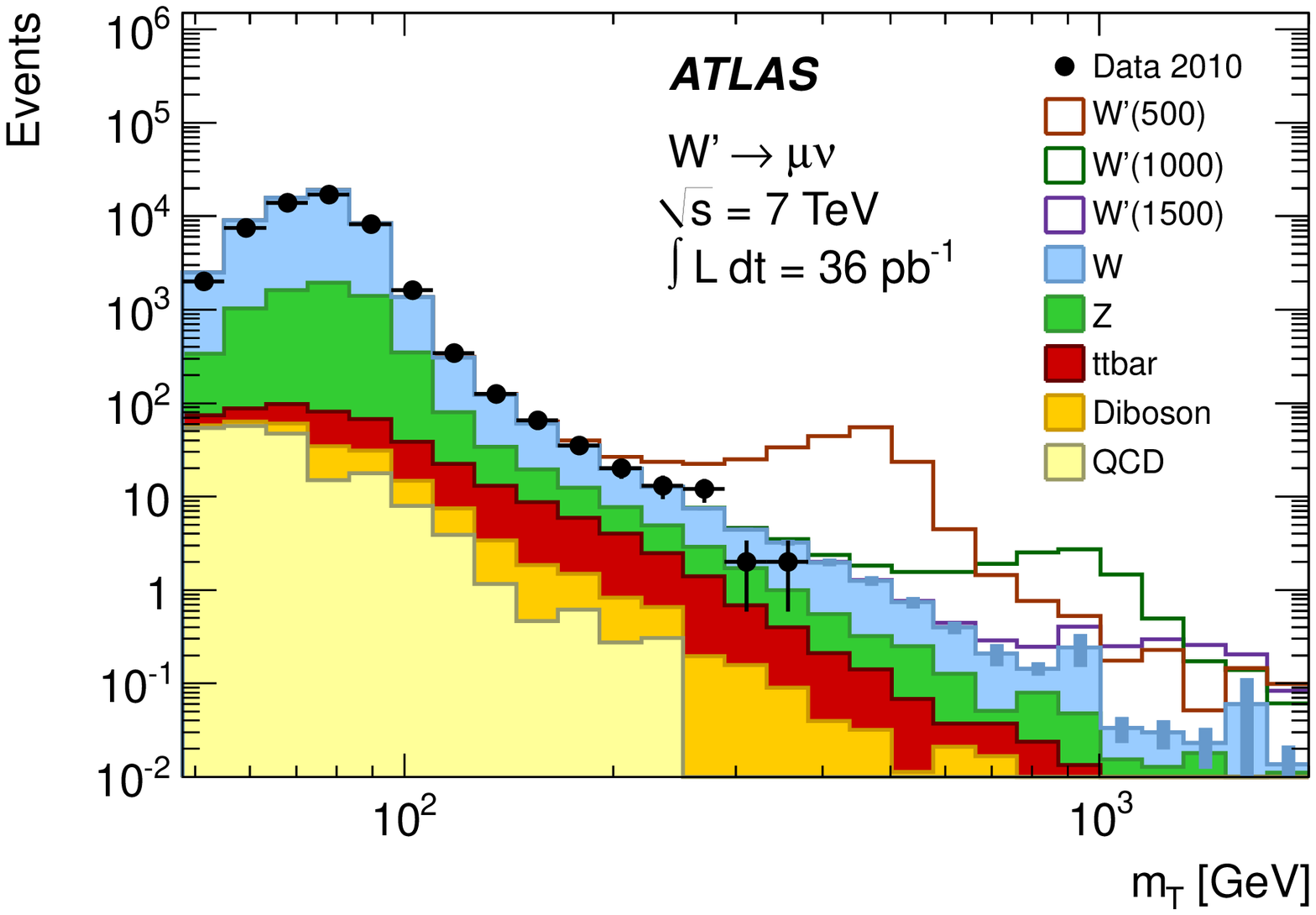}
  \includegraphics[width=0.43\textwidth]{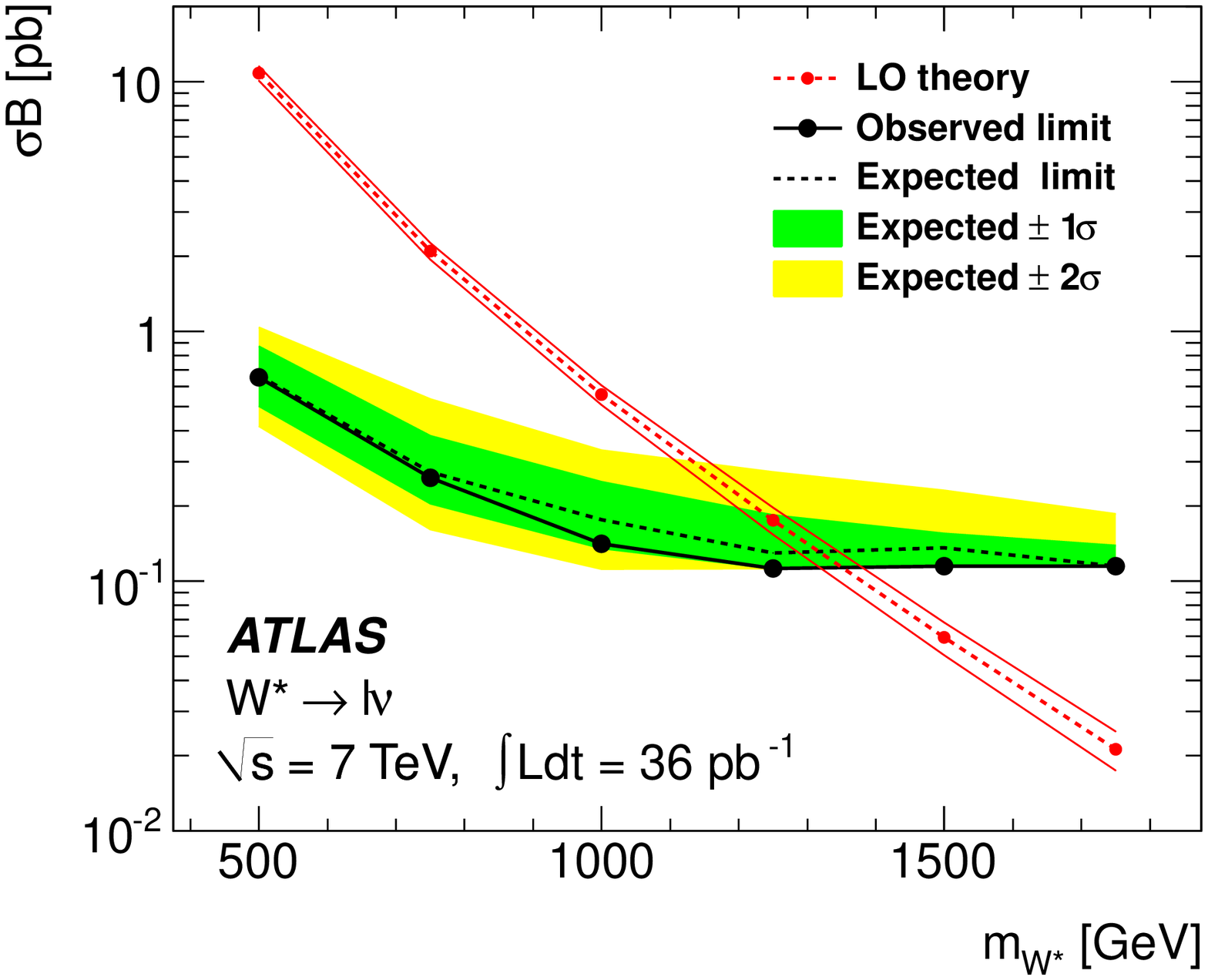}
  \caption{
    Spectra of \pt\ (top), \mettext\ (center) and \mt\ (bottom)
    for the muon channel (left) obtained in
\cite{Aad:2011fe}.
    The points represent the ATLAS data and the
    filled histograms show the stacked backgrounds.
    Open histograms are \wp\ signals.
    The QCD background is estimated from the data.
    Limits at 95\% C.L. for 
    \wstar\ (right) production in the
    decay channels \wpse\ (top), \wpsmu\ (center),
    and their combination (bottom).
    The solid lines show the observed limits with all uncertainties.
    The expected limit is indicated with dashed lines surrounded by
    shaded $1\sigma$ and $2\sigma$ bands.
    Dashed lines show the theory predictions with 
    their uncertainties indicated by solid lines.
  \label{fig:final_mt}
  \label{fig:limits_xbr}
  }
\end{figure*}

      Furthermore, in 2010 the ATLAS collaboration 
      searched for high-mass states,
      such as heavy charged gauge bosons ($W'$, $W^*$),
      decaying to a charged lepton and a neutrino
(see Fig.~\ref{fig:final_mt}).
     The relevant paper ``Search for high-mass states with one
     lepton plus missing transverse momentum
     in proton-proton collisions at $\sqrt{s}= 7$ TeV with the ATLAS detector''
     contains all details of event selection and physical data analysis
\cite{Aad:2011fe}.

      The search for heavy charged resonances inclusively produced at the LHC
      looks more complicated than the search for neutral states
      due to the absence of the second decay particle --- the undetectable neutrino.
      In this case
      the kinematic variable used to identify the \wps\ is the transverse mass
$$
\mt = \sqrt{ 2 \pt \met (1 - \cos \phi_{l\nu})}
$$
      which displays a Jacobian peak that, for $W'\to \ell\nu $,
      falls sharply above the resonance mass.
      Here \pt\ is the lepton transverse momentum,
      \met\ is the magnitude of the missing transverse momentum (\mettext),
      and $\phi_{l\nu}$ is the angle between the \pt\ and \mettext\ vectors.
      In the analysis
\cite{Aad:2011fe}, ``transverse'' refers to the plane
      perpendicular to the colliding beams, ``longitudinal''
      means parallel to the beams, $\theta$ and $\phi$
      are the polar and azimuthal angles with respect to the
      longitudinal direction, and pseudorapidity is defined as
      $\eta = -\ln(\tan(\theta/2))$.

      The main physical results obtained in
\cite{Aad:2011fe} and relevant to our consideration are given in
Fig.~\ref{fig:final_mt}.
     The left panel of Fig.~\ref{fig:final_mt} 
     shows
     the \pt, missing \et, and \mt\ spectra measured in the 
     muon decay channel for the data, for the expected background,
     and for three examples of \wp\ signals at different masses as
     open histograms.
     The $W^*$ boson signals are not shown.
     The QCD background is estimated from the data.
     The signal and other background samples are normalized using
     the integrated luminosity of the data
     and the NNLO (near-NNLO for t-tbar) cross sections. 
     Furthermore, the \xbr\ uncertainties for the $W'$ boson are 
     obtained by varying renormalization and factorization scales and by
     varying PDFs. 
     Only the latter are employed for the $W^*$ boson search.

    One can see from the figures that
    the agreement between the data and the expected background is rather good.
    No excess beyond the Standard Model expectations is observed.
    The lower mass limits expected and obtained
    from these measurements are depicted
    in the right panel of
Fig.~\ref{fig:limits_xbr}. 
    The figure also shows the expected limits and the theoretical \wps\ \xbr\ as a
    function of \mt\ for both channels and their combination.
    The intersection between the central theoretical prediction and the observed limits
    provides the 95\% C.L. lower limit on the mass.
    It was found that the charged chiral $W^*$ boson considered in the paper
    was excluded for masses below 1.350~TeV$/c^2$.
    These are the first direct limits on the $W^*$ boson production.

\section{Dijet signals of the chiral bosons}
    In what follows we will extend the set of possible experimental
    observables of the chiral bosons to the
    rich field of hadron final states, in particular we consider
    peculiarity of the decay of these bosons into two hadronic jets.
    In fact we will draw extra attention to a       novel
    signal of new physics in the dijet data at the hadron colliders.
    It is usually accepted that all exotic models predict that these two jets
    populate the central (pseudo)rapidity region where $y_{1,2}\simeq 0$. 
    Contrariwise, the excited bosons do not contribute into this region,
    but produce an excess of dijet events over the almost flat QCD
    background in $\chi=\exp|y_1-y_2|$ away from this region.

    At the hadron colliders, inclusive dijet production
    has one of the largest cross sections
    and allows data-driven background estimation at the early
    stage of the collider operation.
    The feature can be used to search for a signal of new physics
    in the very early data.
    In particular, a possible bump in the dijet invariant mass
    spectrum would indicate the presence of
    a resonance decaying into two energetic partons.
    Nevertheless, we could say nothing about its nature, because
    this bump 
    stems from the Breit--Wigner propagator form,
    which is characteristic of any type of resonance regardless
    of its other properties, like spin, internal quantum number, etc.
    Therefore, other observables are necessary in order to confirm
    the bump and to reveal the resonance properties.

    As in the lepton case (discussed above), this
    kind of observable could be the dijet distribution over
    the polar angle $\theta$, which is an angle between the
    axis of the jet pair and the beam direction in the dijet rest frame.
    This distribution is directly sensitive to the resonance spin and
    the dynamics of the underlying process
\cite{Collins:1977iv}.

    While the QCD processes are dominated by $t$-channel gluon exchanges,
    which lead to the Rutherford-like dijet distribution $1/(1-\cos\theta)^2$,
    exotic physics processes proceed mainly through the $s$-channel,
    where the spin of the resonance uniquely defines the angular distribution.
    For high-mass resonances and practically massless partons it is
    convenient to use the helicity formalism since the helicity is a
    good quantum number for massless particles.

    In fact, in the center-of-momentum frame of a particle with spin 
    $s$ and helicity $\lambda$ ($-s\leq\lambda\leq s$) decaying into 
    two massless particles with helicities $\lambda_1$ and $\lambda_2$
    the angular distribution of the outgoing particle can be written as
\cite{Haber:1994pe}
\begin{equation}\label{ds/dcos}
    \frac{{\rm d}\Gamma_s}{{\rm d}\cos\theta\;{\rm d}\phi}=
    \frac{1}{64\pi^2 M}\,
\left\vert\,    
\sqrt{\frac{2s+1}{4\pi}}\,e^{i(\lambda-\delta)\phi}\,
    d^s_{\lambda\delta}(\theta)\,{\cal M}^s_{\lambda_1\lambda_2}
\right\vert^2,
\end{equation}
    where
    $\delta\equiv\lambda_1-\lambda_2$ with $-s\leq\delta\leq s$.
    The reduced decay amplitude ${\cal M}^s_{\lambda_1\lambda_2}$ is
    only a function of $s$ and the helicities of the outgoing particles.
    It does not dependent on the azimuthal $\phi$ and polar $\theta$ angles.
    The $\theta$ dependence is concentrated only in the well-known
    $d$-functions $d^s_{\lambda\delta}(\theta)$.
    Furthermore, the absolute value of the dijet rapidity difference 
    is related to the polar scattering angle $\theta$ with respect to 
    the beam axis by the formula 
    $\Delta y\equiv |y_1-y_2|=\ln[(1+|\cos\theta|)/(1-|\cos\theta|)]\ge 0$ 
    and is invariant under boosts along the beam direction. 
    The choice of the other variable 
$\chi\equiv\exp(\Delta y)=(1+|\cos\theta|)/(1-|\cos\theta|)\ge 1$ 
    is motivated by the fact  that the distribution 
    of the Rutherford scattering is flat in this variable.
    These variables allow
    systematic consideration of angular decay distributions
    of resonances with different spins and different interactions with partons.

\begin{figure}[!ht]
\epsfig{file=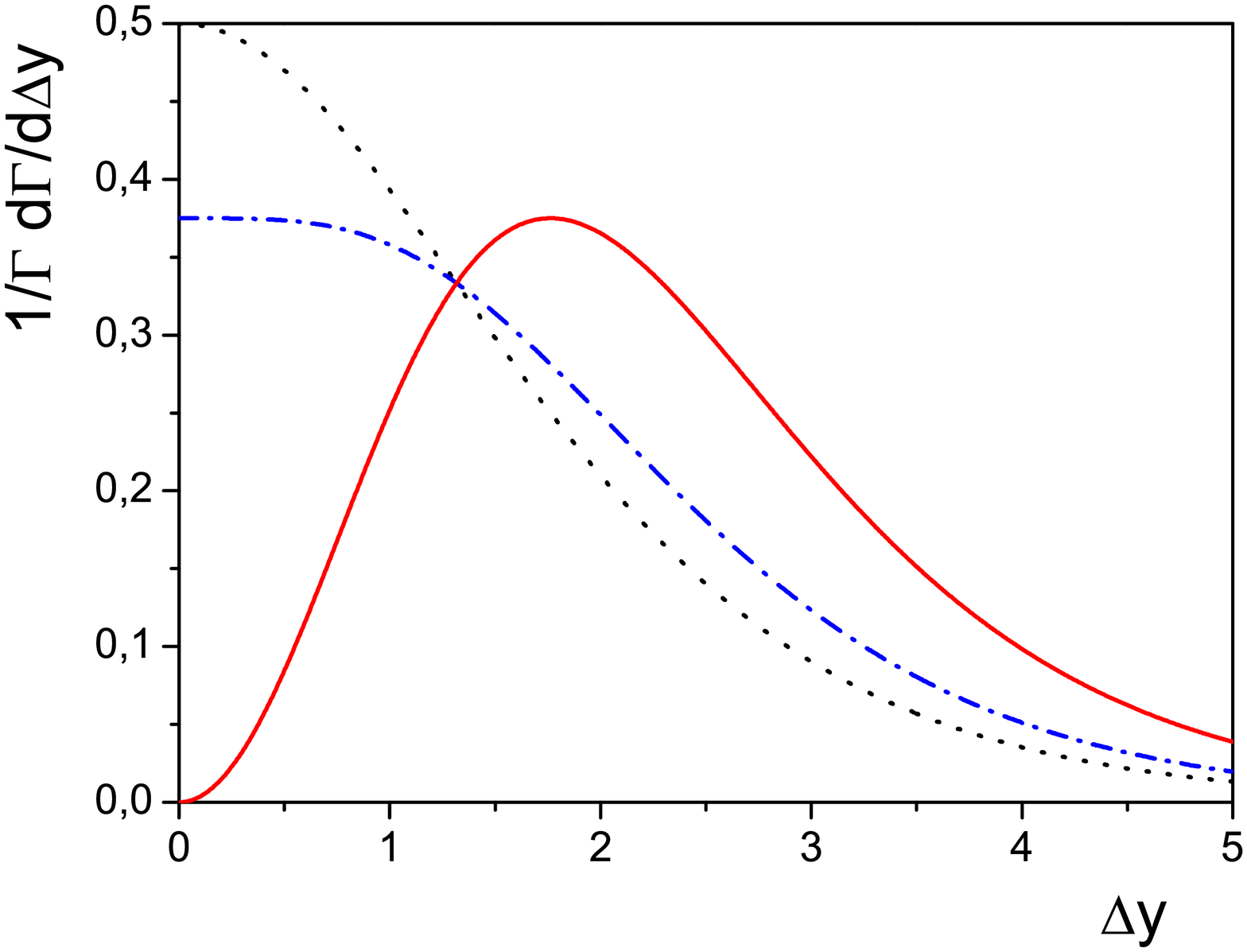,width=0.5\textwidth}%
\epsfig{file=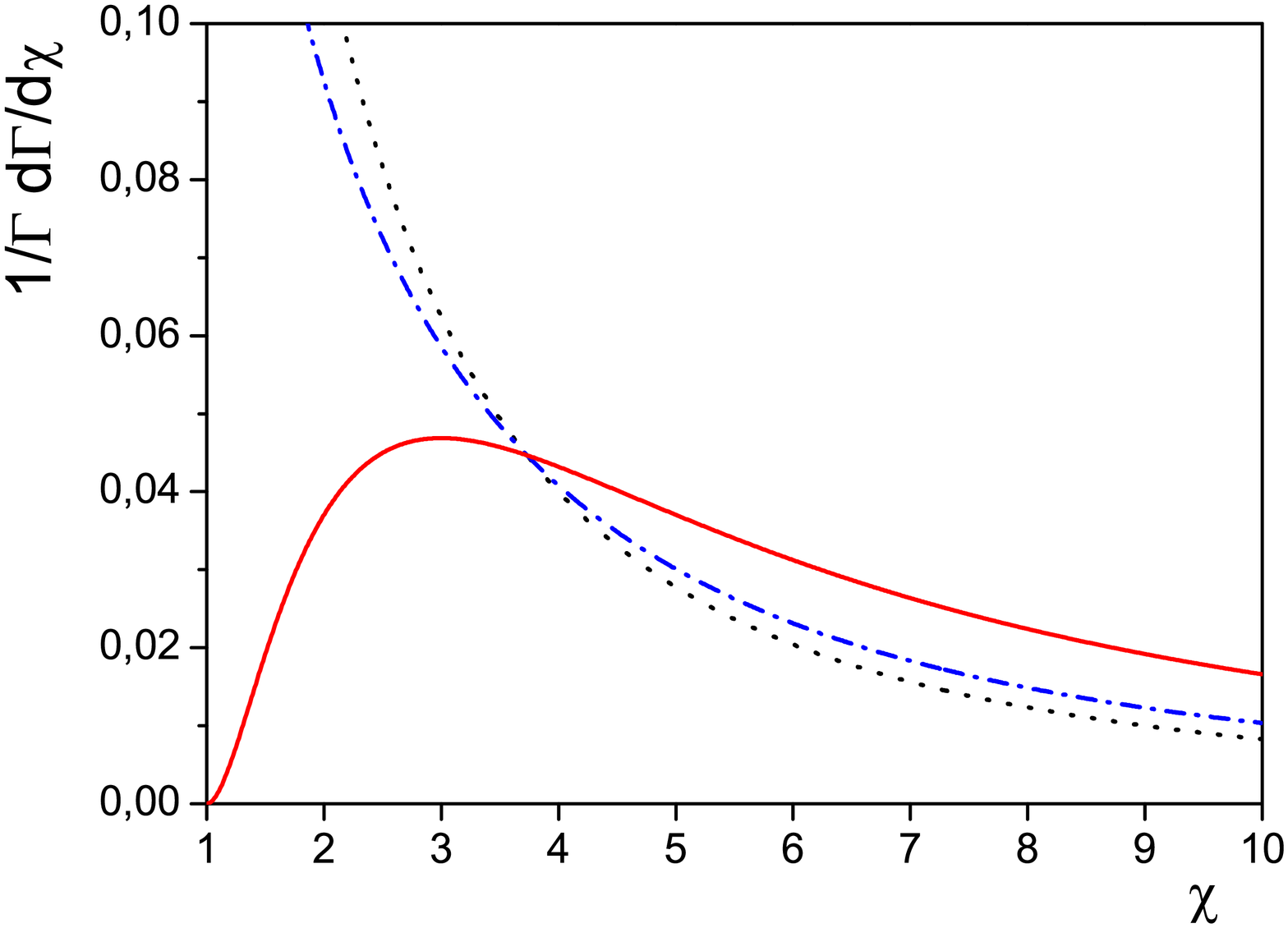,width=0.5\textwidth}
\caption{\label{fig:angular}
      The normalized angular dijet
      distributions as functions of the absolute value of the rapidity
      difference (left) and the $\chi$ variable (right) for the scalar
      (or/and spin-1/2) bosons, the gauge bosons with the minimal coupling and
      the excited bosons are shown by the dotted,
      dash-dotted and solid curves, respectively.
From \cite{Chizhov:2010jg}.}
\end{figure}

    The simplest case of the
    resonance production of a (pseudo)scalar particle $h$ with spin 0 in
    the $s$-channel leads to a uniform decay distribution in 
    the scattering angle
\begin{equation}\label{G0}
    \frac{{\rm d} \Gamma_0(h\to q\bar{q})}
    {{\rm d} \cos\theta} \propto \vert d^{\,0}_{00}\vert\,^2\sim 1.
\end{equation}
     The spin-1/2 fermion resonance, like an excited quark $q^*$, leads
     to asymmetric decay distributions for the given spin parton
     configurations
\begin{equation}\label{G12}
       \frac{{\rm d} \Gamma_{1/2}(q^*\to qg)}
                     {{\rm d} \cos\theta}
\propto  \vert d^{1/2}_{1/2,\pm 1/2}\vert\,^2\sim 1 \pm \cos\theta
.
\end{equation} 
      However, the choice of the variables which depend on the absolute
      value of $\cos\theta$ cancels out the apparent dependence on
      $\cos\theta$.
      In other words, both distributions (\ref{G12}) 
      for dijet events look like uniform distributions in
      $\Delta y$ and $\chi$
      According to the simple formula
\begin{equation}\label{trans}
    \frac{{\rm d}\Gamma}{{\rm d}(\Delta y/\chi)}=
    \frac{{\rm d}\cos\theta}{{\rm d}(\Delta y/\chi)}\;
    \frac{{\rm d}\Gamma}{{\rm d}\cos\theta}\,,
\end{equation}
       the uniform distribution leads to kinematical peaks at the 
       small values $\Delta y=0$ (the dotted curve in the left panel of
Fig.~\ref{fig:angular})
       and $\chi=1$ (the dotted curve in the right panel of
Fig.~\ref{fig:angular})
\begin{equation} 
    \frac{{\rm d}\Gamma_0}{{\rm d}\Delta y}\propto
    \frac{{\rm e}^{\Delta y}}{({\rm e}^{\Delta y}+1)^2}
\qquad {\rm and} \qquad \label{chi0}
    \frac{{\rm d}\Gamma_0}{{\rm d}\chi}\propto
    \frac{1}{(\chi+1)^2}.
\end{equation}

      There are two different possibilities for spin-1 resonances.
      The gauge bosons, which are associated with additional $U(1)'$ gauge
      symmetry (or transform under the adjoint representation of the extra
      gauge group), are generally called $Z'(\boldsymbol{W}')$ particles.
      They have minimal gauge interactions with the known light fermions
(see eqs. (\ref{Z'ed}) and
(\ref{W'})),
    which preserve the fermion chiralities and possess maximal
    helicities $\lambda=\pm 1$.
    At a symmetric $pp$ collider, like the LHC,
    such interactions lead to the symmetric angular distribution
(\ref{GLL})
    of the decay products over the polar angle $\theta$:
\begin{equation}\label{GLL11}
    \frac{{\rm d} \Gamma_1(Z'\to q\bar{q})}
    {{\rm d} \cos\theta} \propto
    \vert d^1_{11}\vert^2+\vert d^1_{-11}\vert^2 \sim
    1+\cos^2\theta \, .
\end{equation}
       Similar to the uniform distribution
(\ref{G0}), this one also 
       leads to kinematical peaks at small values of $\Delta y=0$
       (the dash-dotted curve in the left panel of
Fig.~\ref{fig:angular})
        and $\chi=1$  (the dash-dotted curve in the right panel of
Fig.~\ref{fig:angular})
\begin{equation}
\label{dy1prime}
    \frac{{\rm d}\Gamma'_1}{{\rm d}\Delta y}\propto
    \frac{{\rm e}^{\Delta y}({\rm e}^{2\Delta y}+1)}{({\rm e}^{\Delta y}+1)^4}
\qquad {\rm and } \qquad 
    \frac{{\rm d}\Gamma'_1}{{\rm d}\chi}\propto
    \frac{\chi^2+1}{(\chi+1)^4}.
\end{equation}

\begin{figure}[h]
\epsfig{file=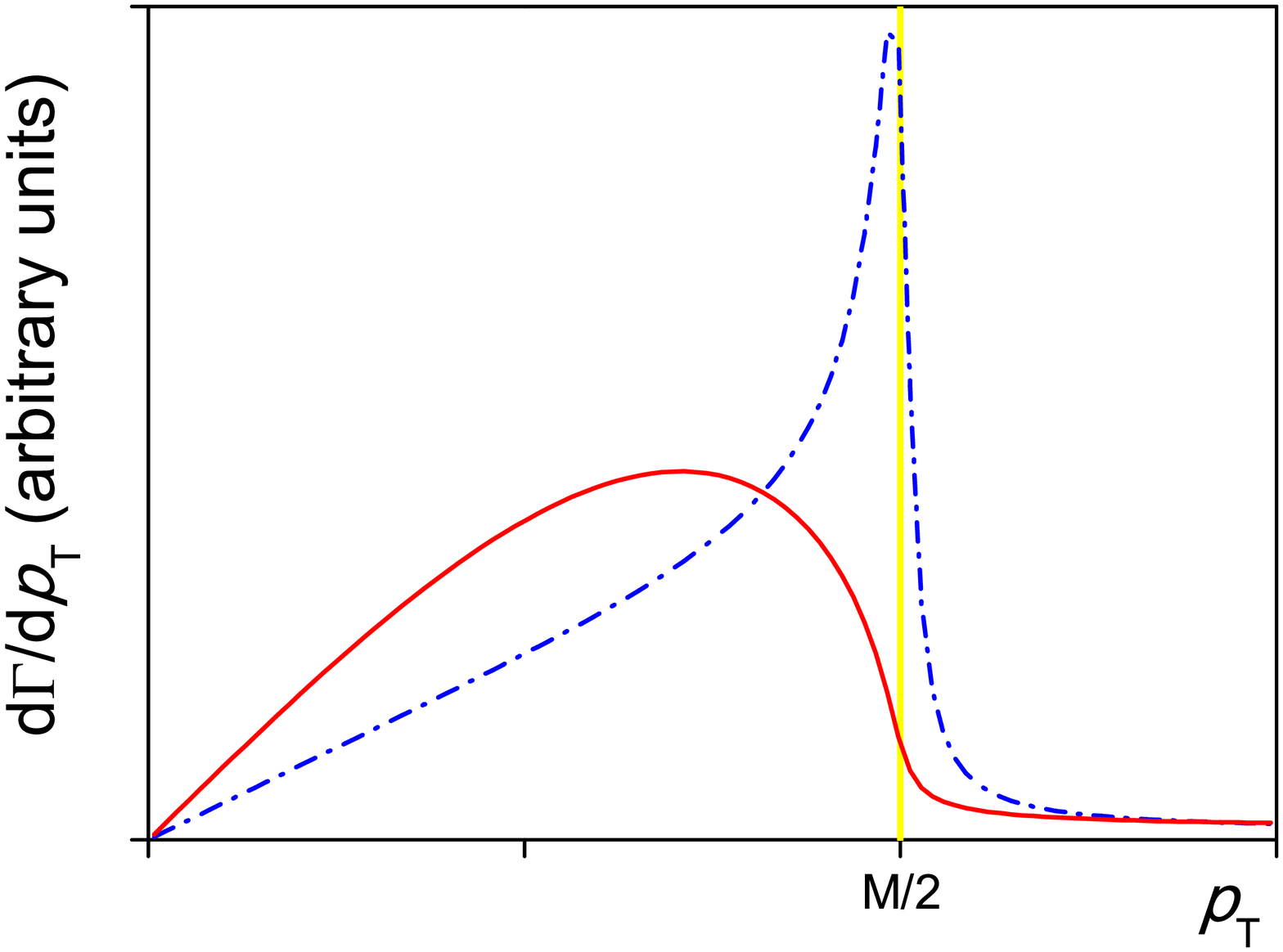,width=0.49\textwidth}%
\epsfig{file=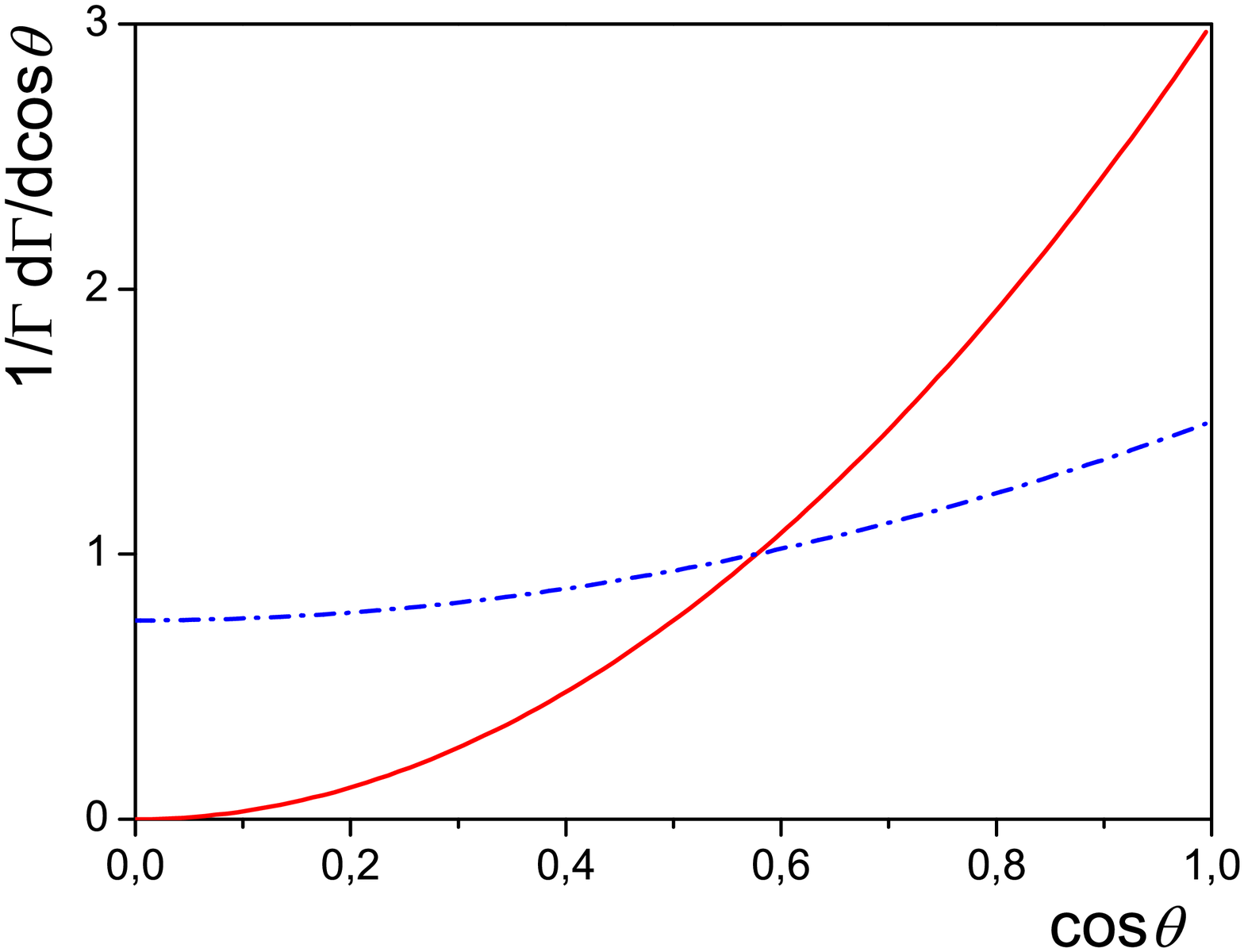,width=0.49\textwidth}
\caption{\label{fig:pt}
     The final parton transverse momentum (left)
     and the angular (right) distributions from the decay of the gauge 
     (dash-dotted curves) and excited (solid curves) bosons.
From \cite{Chizhov:2010jg}.
}
\end{figure}

    Another possibility is the resonance production and decay of new
    longitudinal spin-1 bosons with helicity $\lambda=0$.
    These bosons arise in many extensions
\cite{Chizhov:2009fc} of SM which solve the Hierarchy problem.
    They are transformed as doublets $(Z^*\;W^*)$ under the fundamental
    representation of the SM $SU(2)_W$ group like the SM Higgs boson.
    They are above-mentioned extra chiral bosons
(see (\ref{master})).
    While the $Z'$ bosons with helicities $\lambda=\pm 1$ are produced
    in left(right)-handed quark and right(left)-handed antiquark fusion,
    the longitudinal $Z^*$ bosons can be produced through the anomalous
    chiral couplings with the ordinary light fermions
    in left-handed or right-handed quark-antiquark fusion
\cite{Chizhov:2008tp}.
    As already noted before, these anomalous couplings lead 
    to a different angular distribution of the resonance decay
\begin{equation}\label{GLR11}
    \frac{{\rm d} \Gamma^*_1(Z^*\to q\bar{q})}
    {{\rm d} \cos\theta} \propto
    \vert d^1_{00}\vert\,^2\sim\cos^2\theta.
\end{equation}
        As has already been noted for the dilepton case the
	absence of the constant term in
(\ref{GLR11}) results in novel experimental signatures.
    First of all, the uniform distribution (\ref{G0}) for scalar and
    spin-1/2 particles and the distribution (\ref{GLL11}) for gauge vector
    bosons with minimal coupling include a nonzero constant term, which
    leads to a kinematic singularity in the transverse momentum
    distribution of the final parton (as for charged leptons, see
(\ref{1/cos}) and (\ref{narrow})).
     After smearing of the resonance finite width the singularity is
     transformed into the well-known Jacobian peak (the dash-dotted curve
     in the left panel of
Fig.~\ref{fig:pt}).
      The analytic expression of the $p_{\rm T}$ distribution 
      describing the Jacobian peak with finite width can be found in
\cite{Barger:1983wf}.
     Using the same method one can derive an analogous distribution for
     the excited bosons (the solid curve in the left panel of
Fig.~\ref{fig:pt}).
\begin{equation}\label{pTanal}
    \frac{{\rm d}\Gamma^*_1}{{\rm d}p_{\rm T}}\propto
    p_{\rm T}\sqrt{\sqrt{\left(4p^2_{\rm T}-M^2\right)^2+\Gamma^2 M^2}-4p^2_{\rm
    T}+M^2}.
\end{equation}
     In contrast to the previous case, the pole in the decay distribution
     of the excited bosons is canceled out and the final parton $p_{\rm
       T}$ distribution has a broad smooth hump
\cite{Chizhov:2006nw} with a maximum
    at $p_{\rm T}=\sqrt{(M^2+\Gamma^2)/8}\simeq M/\sqrt{8}$ below the
    kinematic endpoint $p_{\rm T}=M/2$ instead of a sharp Jacobian peak, 
    which obscures their experimental identification as resonances.
    Therefore, the transverse jet momentum is not the appropriate
    variable for the excited boson search.

       Another striking feature of the distribution (\ref{GLR11}) is the
       forbidden decay direction perpendicular to the boost of the excited
       boson in the rest frame of the latter (the Collins--Soper frame
\cite{Collins:1977iv}).
       It leads to a profound dip at $\cos\theta=0$ in the Collins--Soper frame
\cite{Chizhov:2008tp} in comparison with the
     gauge boson distribution (the right panel in
Fig.~\ref{fig:pt}).
     Similar dips also occur at the small values $\Delta y=0$
\cite{Chizhov:2010hh} (the solid curve in the left panel of
Fig.~\ref{fig:angular})
     and $\chi=1$ (the solid curve in the right panel of
Fig.~\ref{fig:angular})
\begin{equation}
\label{dy1star}
    \frac{{\rm d}\Gamma^*_1}{{\rm d}\Delta y}\propto
    \frac{{\rm e}^{\Delta y}({\rm e}^{\Delta y}-1)^2}{({\rm e}^{\Delta y}+1)^4}
\qquad {\rm and } \qquad
    \frac{{\rm d}\Gamma^*_1}{{\rm d}\chi}\propto
    \frac{(\chi-1)^2}{(\chi+1)^4}.
\end{equation}
      It can be seen from Fig.~\ref{fig:angular} that the excited bosons
      have a unique signature in the angular distributions.
      They manifest themselves through the absolute minima 
      at the small values $\Delta y=0$ and $\chi=1$ 
      and absolute maxima right away from the origin.
      So, the rapidity difference distribution reaches the absolute
      maximum at $\Delta y=\ln(3+\sqrt{8})\approx 1.76$ and at $\chi=3$
      for the angular distribution in the dijet variable $\chi$.

\begin{figure}[th]
\epsfig{file=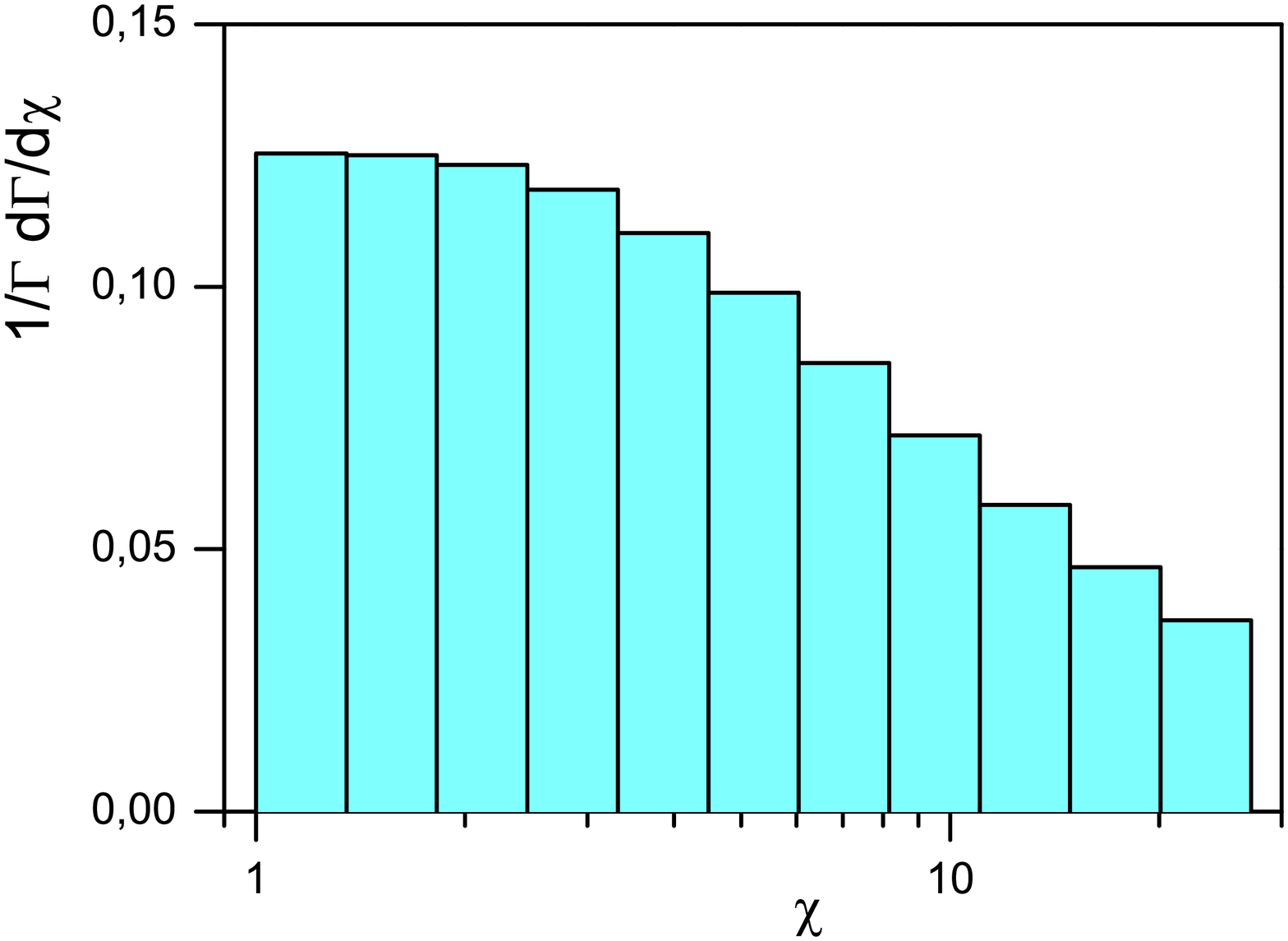,width=0.5\textwidth}%
\epsfig{file=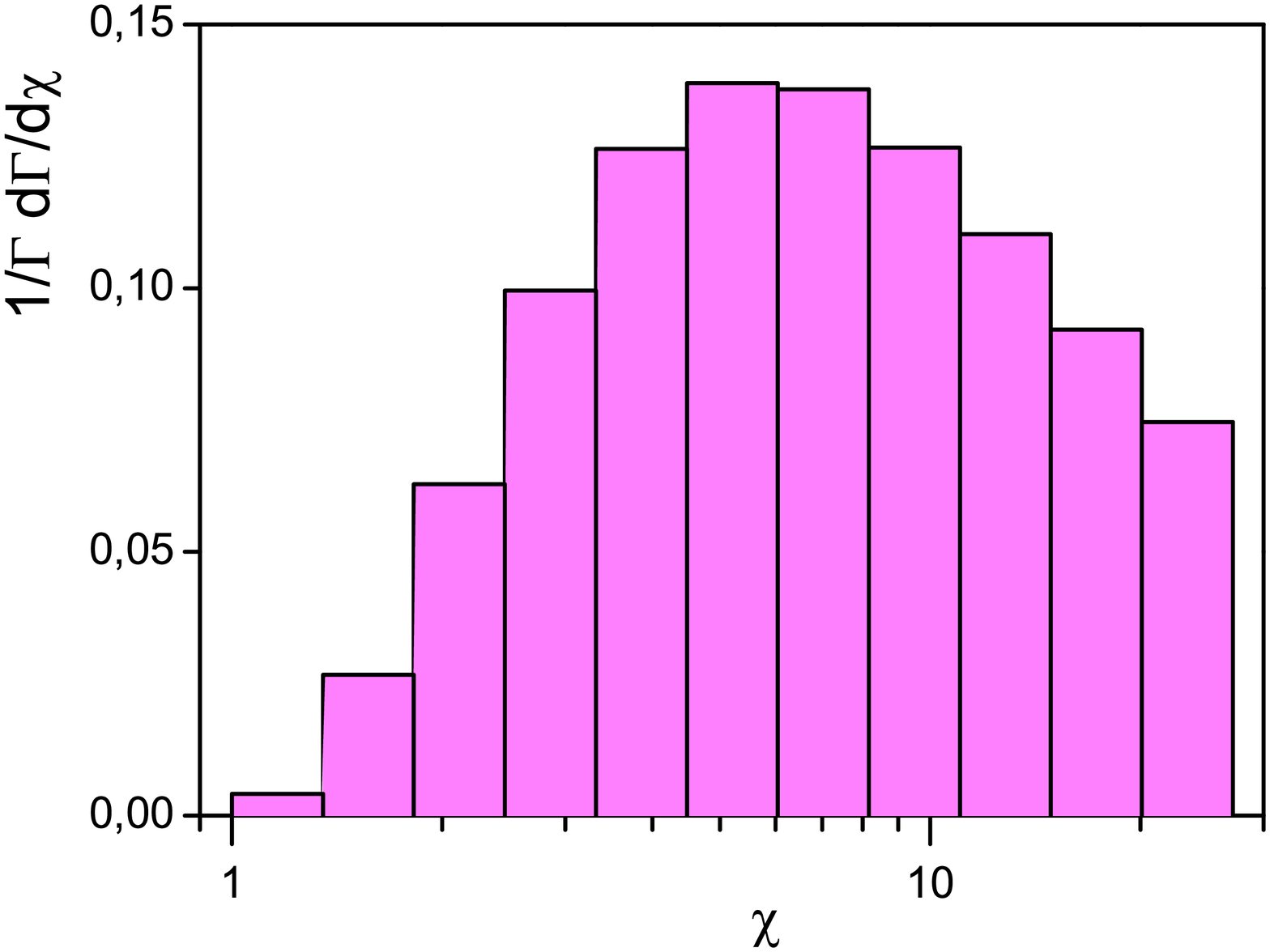,width=0.5\textwidth}
\caption{\label{fig:chi}
   The normalized histograms of $\chi$-spectra
   for the gauge bosons with the minimal coupling (left) and for the
   excited bosons (right).
From \cite{Chizhov:2010jg}.}
\end{figure}

\smallskip 
      These features will be considered below in more detail.
      In order to have more practical analysis,
      it is convenient to use equidistant binning in $\log\chi$
\cite{Collaboration:2010eza},
      which corresponds to periodic cell granularity of the 
      calorimeter in $\eta$.
      In this case the smooth $\chi$-spectra
 (see eqs. (\ref{dy1prime}) and (\ref{dy1star}))
      are transformed into histograms with the maximum in the lowest bin for the
      gauge bosons with the minimal coupling and with the maximum in the bin
      containing the value $\chi=3+\sqrt{8}\approx 5.8$ for the excited bosons
(Fig.~\ref{fig:chi}).

      Using distributions in (pseudo)rapidity and $\chi$ one can construct
      two useful ratios of numbers of events $N$
      measured under specified experimental
      constraints schematically given below in brackets.
      The first is the wide-angle to small-angle ratio
\begin{equation}\label{wide-small}
    R_\chi(a,b)=\frac{N(1<\chi<a)}{N(a<\chi<b)}
\end{equation}
      and the second is centrality or the $\eta$-ratio of both jets
\begin{equation}\label{centrality}
    R_\eta(a,b)=\frac{N(|\eta_{1,2}|<a)}{N(a<|\eta_{1,2}|<b)}.
\end{equation}

      The ratios are less affected by the systematic uncertainties
      and can be used for searching for new physics in dijet data.
      To understand how they work, let us suppose that one has found some bump
      in the experimental dijet invariant mass distribution.
      Then we can compare the angular distributions for ``on-peak'' events
      (or events comprising 
      the bump) and ``off-peak'' events
      (or events far away from the bump), using the aforementioned ratios.
      At the moment we ignore the complications of experimental
      separation of ``on-peak'' events from the ``off-peak'' ones.
      Since the QCD background is dominated by the Rutherford-like
      distribution, we can consider, as an approximation, a simple case
      where the QCD dijet $\chi$-distribution is flat.
      It means that for the selected equal kinematical regions $b-a=a-1$ ($b=2a-1$),
      the ratio $R_\chi$ for the ``off-peak'' events should be approximately
      one and does not depend on the dijet mass.
      When the ``on-peak'' events originate from the
      new physics and have the angular distribution different from the
      one predicted in QCD, the ratio $R_\chi$ should deviate from one.
      Due to an excess at small $\chi$ values irrespectiveof the  
      maximal $\chi$-value (in our case equals to $a$)
      one has $R_\chi>1$ for all known exotic models,
      except the excited bosons, when one expects $R_\chi<1$.

      In order to emphasize the effect of the excited bosons 
      (to increase sensitivity to these bosons), 
      we need to choose a value of $a$,
      which makes the ratio as small as possible
\begin{equation}\label{minChi}
R_\chi(a,2a-1)=
\frac{N_{\rm QCD}+N_{\rm new}(<a)}
     {N_{\rm QCD}+N_{\rm new}(>a)}
\approx 1+
\frac{N_{\rm new}(<a)-N_{\rm new}(>a)}
     {N_{\rm QCD}}<1.
\end{equation}
        Here $N_{\rm new}(<a)$ and $N_{\rm new}(>a)$
	denotes the number of events generated
	by the new physics in the regions $1<\chi<a$
	and $a<\chi<2a-1$, respectively.
	Simple integration gives the QCD contribution
    $N_{\rm QCD}(1<\chi<a)    \propto \int^{a}_{1} d\chi
    =N_{\rm QCD}(a<\chi<2a-1) \propto \int^{2a-1}_{a} d\chi = (a-1)$.
    Simialr integrations of the $\chi$-distribution
(\ref{dy1star}) in $1<\chi<a$ and $a<\chi<2a-1$ give 
    $N_{\rm new}(<a) \propto (a-1)^3/(a+1)^3$ and
    $N_{\rm new}(>a) \propto (a-1)^3/a^3-N_{\rm new}(<a)$, respectively.
    Therefore,
    due to the monotonic increase of the distribution
(\ref{dy1star}) to the maximum at $\chi=3$
    it is possible to reach the minimal value for
(\ref{minChi})  with the parameters $a\approx 1.87$ and $b\approx 2.74$.
    In our ``ideal'' case these parameters are optimal for the search for
    excited bosons with the ratio
(\ref{wide-small}).
    The larger value $a\ge
    1/(\sqrt[3]{2}-1)\approx 3.85$ will lead to a compensation of the
    contributions from the low and high $\chi$-parts
    ($N_{\rm new}(<a) \approx N_{\rm new}(>a)$),
    and $R_\chi \ge 1$ again.

         For the QCD generated dijets the centrality ratio $R_\eta$
(\ref{centrality}) is also almost constant 
         and should not depend on the dijet invariant mass
         when the parameters $a$ and $b$ are fixed.
	 When a dijet new-physics signal takes place,
      this ratio could deviate from its constant value.
      Signal events for almost all exotic models
      are expected in the central (pseudo)rapidity region.
      Therefore, one could see a bump in the $R_\eta$
      distribution as a function of the dijet mass.
      Contrariwise, the signal from the excited bosons
      could lead to a novel signature: 
      instead of the bump one will have a dip
      in the distribution at the resonance mass.

\begin{figure}[th]
\vspace*{-0.6cm}
\hspace{-0.5cm}\epsfig{file=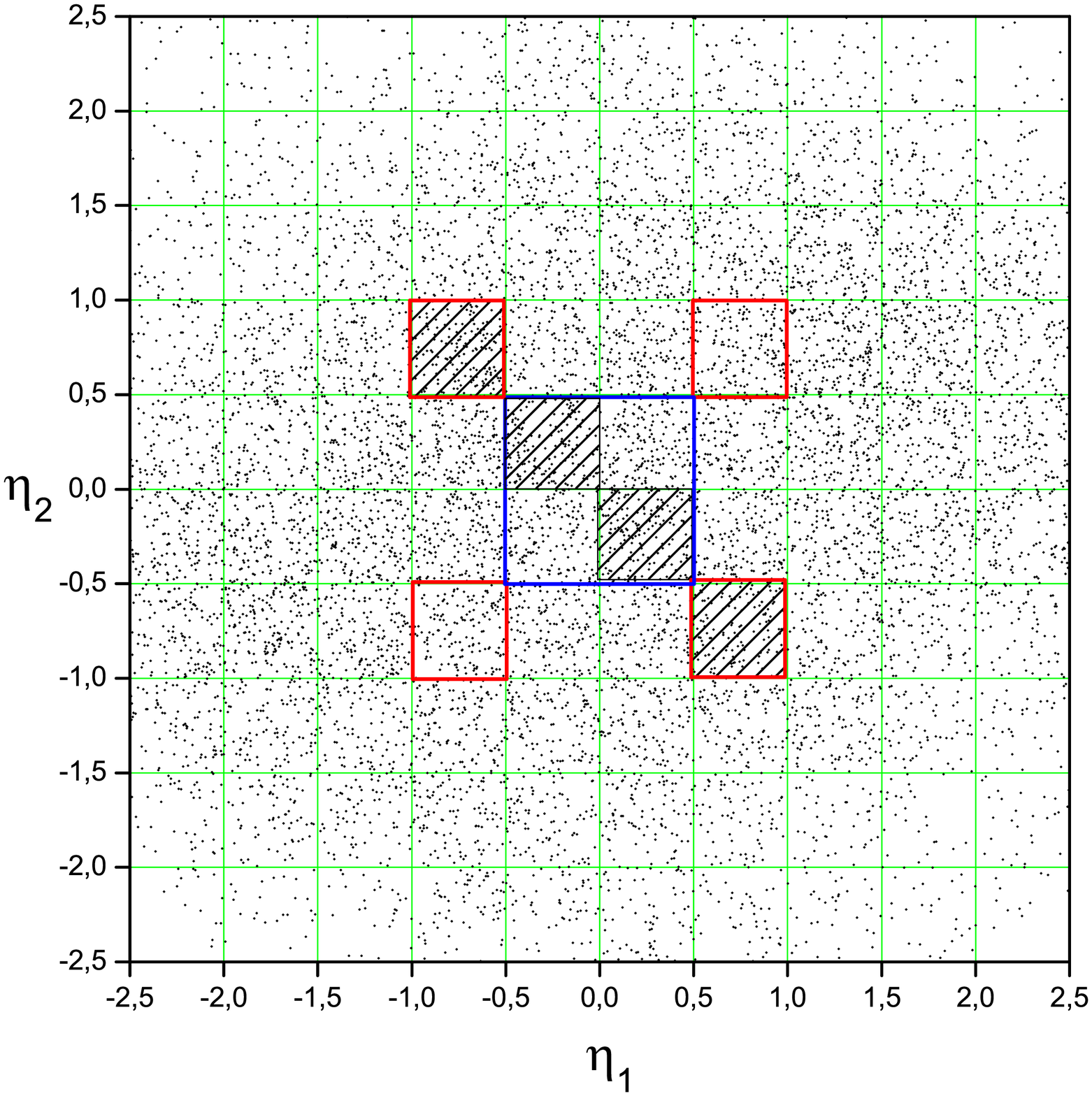,width=0.55\textwidth}
\hspace{-1.5cm}\epsfig{file=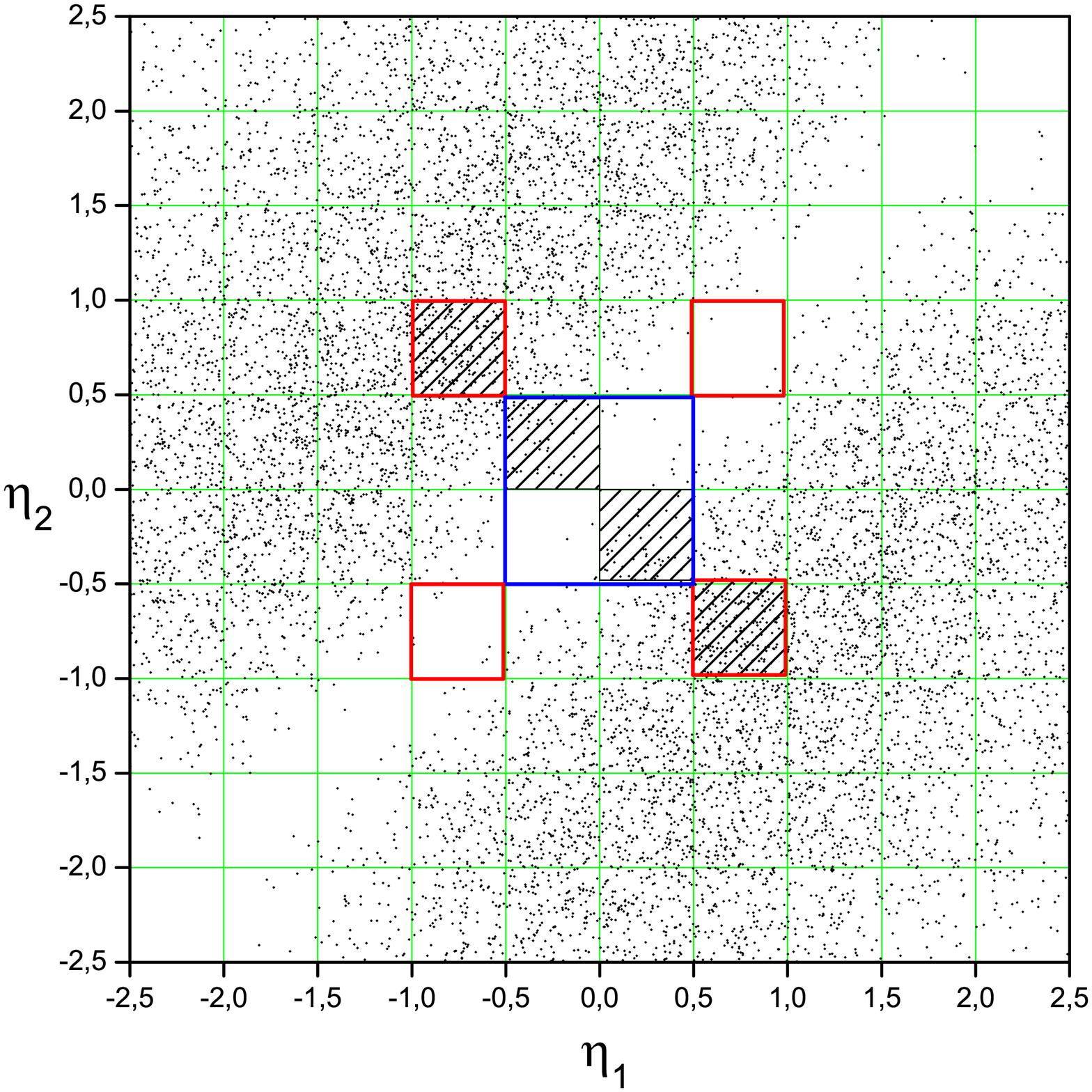,width=0.55\textwidth}
\vspace*{-0.6cm}
\caption{\label{fig:eta2dim}
       The two dimensional pseudorapidity distributions
       (scatter plots) of dijet events for the gauge bosons
       with the minimal coupling (left) and for the excited bosons (right).
       The central region $|\eta_{1,2}|<0.5$ and outer regions
       $0.5<|\eta_{1,2}|<1.0$ are depicted.
       The hatched squared regions
       correspond to selected events with opposite pseudorapidity signs.
From \cite{Chizhov:2010jg}.
}
\end{figure}

      To investigate quantitatively this possibility
      we have again used the CompHEP package
\cite{Boos:2004kh,Pukhov:1999gg}, which was extended with the excited bosons model of
\cite{Chizhov:2010ry}.
         In particular, two dimensional pseudorapidity distributions
     were generated
     for ``$\,2\to 2\,$'' processes
     proceeding through the gauge $W'$ and excited $W^*$ boson resonances
     with the same mass 550 GeV$/c^2$ at $\sqrt{s}=7$~TeV $pp$ collider
(Fig.~\ref{fig:eta2dim}).
     The CTEQ6L parton distribution functions were used. For both final
     jets we impose cuts on the pseudorapidity $|\eta|<2.5$ and the
     transverse momentum $p_{\rm T} > 30$~GeV/$c$.
           To minimize the potential differences in jet response
       and efficiency of jet registration between the
       {\em inner} and {\em outer} dijet events one can choose the central
       region of the calorimeter $|\eta|<1$.
       Using calculations given in the scatter plots
(Fig.~\ref{fig:eta2dim}) one can estimate the centrality ratios
(\ref{centrality}) for the gauge and excited bosons
\begin{equation}\label{etaRprime}
    R'_\eta(0.5,1.0)\simeq 1.08 \qquad {\rm and} \qquad
    R^*_\eta(0.5,1.0)\simeq 0.29.
\end{equation}
      The dramatic difference between these two numbers is clearly seen, 
      it should lead to the corresponding experimental signature.
      Since the QCD ratio $R^{\rm QCD}_\eta(0.5,1.0)\simeq 0.6$
      is located {\em right between}\/
      the numbers, the gauge bosons with the minimal coupling will
      lead to an increase in the QCD ratio at the
      resonance mass, while the excited bosons should decrease the ratio.
      It is interesting to notice that a hint of this type 
      of the novel signature can be seen in the low-statistics distribution of
      the $\eta$-ratio versus the dijet invariant mass in the ATLAS data
      in approximately the same mass range,
      $450~{\rm GeV}<M_{jj}<600$~GeV, as for the resonance bump in the
      dijet events~\cite{:2010bc}.

\begin{figure}[th]
\epsfig{file=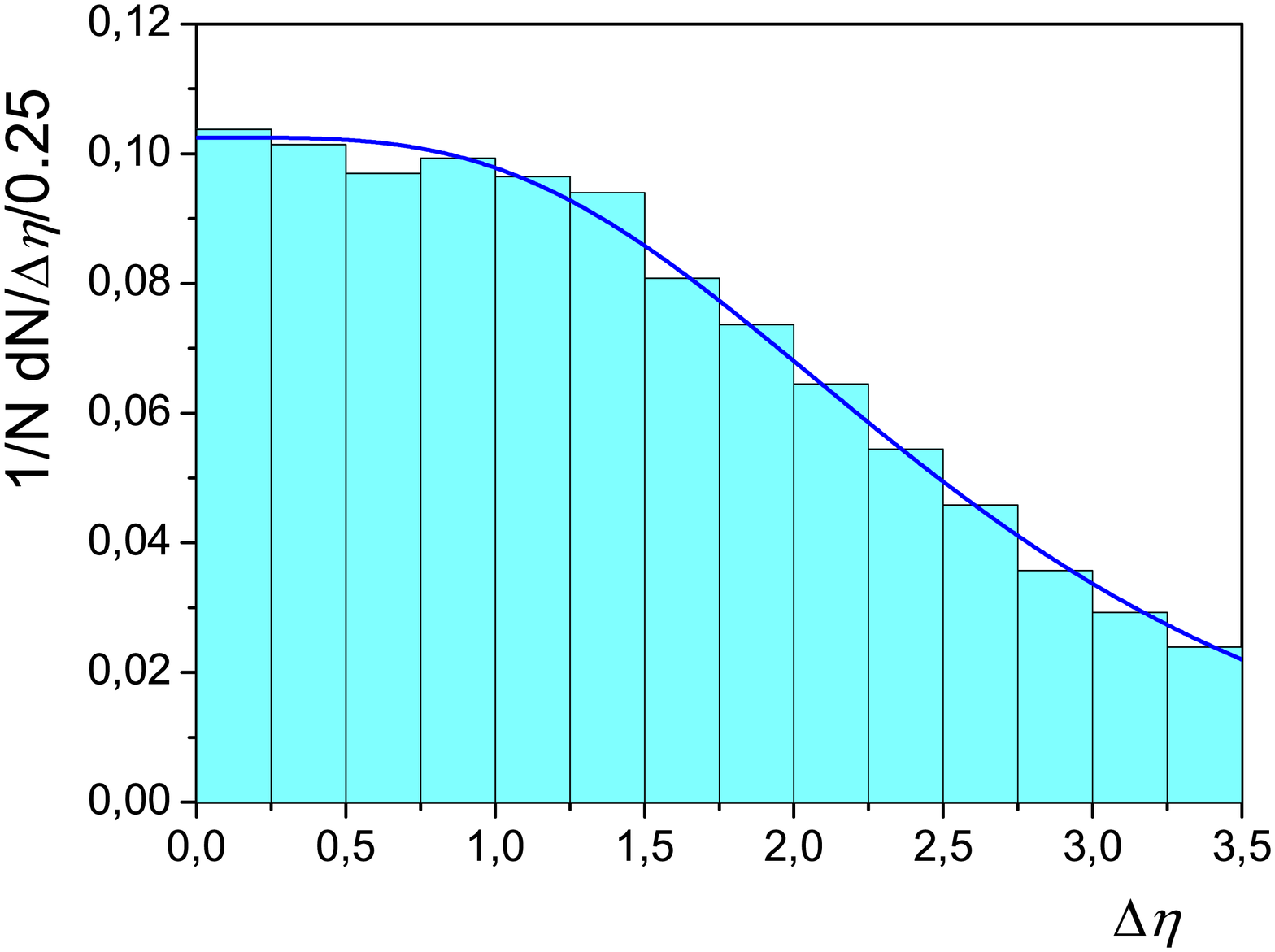,width=0.5\textwidth}%
\epsfig{file=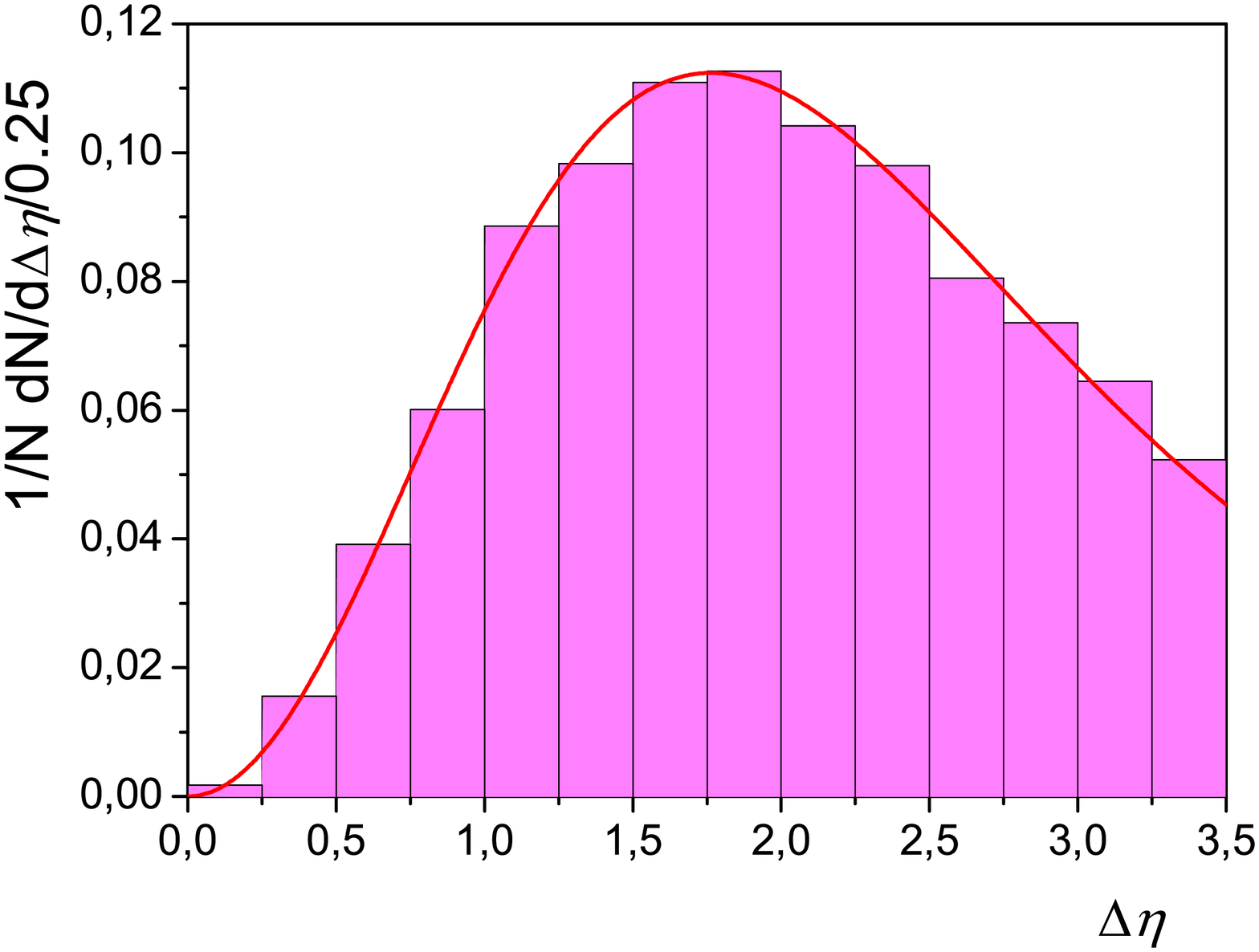,width=0.5\textwidth}
\caption{\label{fig:eta1}
    The normalized histograms of the
    $\Delta\eta\,$-spectra for the gauge bosons with the minimal
    coupling (left) and for the excited bosons (right). The solid curves
    correspond to the theoretical formulas
(\ref{dy1prime}) and
(\ref{dy1star}).
From \cite{Chizhov:2010jg}.
}\end{figure}

      Unfortunately, one should stress
      that the extensions of the ``signal region'' up to
      $|\eta|<1.3$ and the central region up to $|\eta|<0.7$ do not change
      drastically the QCD ratio $R^{\rm QCD}_\eta(0.7,1.3)\simeq 0.55$,
      but dilute the signal from the excited bosons since
      $R^*_\eta(0.7,1.3)\simeq 0.68$.
      In
      order to increase the sensitivity to the excited bosons 
      one can consider the centrality ratio only for the dijet 
      events with the opposite   pseudorapidities
$\left.R_\eta\right|_{(\eta_1\cdot\eta_2)\le 0}={\cal R}_\eta$
      (the hatched regions in Fig.~\ref{fig:eta2dim}).
      In this case the difference between ${\cal R}'$ and ${\cal R}^*$ 
      increases
\begin{equation}\label{etaR0prime}
    {\cal R}'_\eta(0.5,1.0)\simeq 1.12
\qquad {\rm and} \qquad
    {\cal R}^*_\eta(0.5,1.0)\simeq 0.25,
\end{equation} 
      but we lose half of the statistics.
      Therefore, it is convenient to consider the distribution
      in 
      $\Delta\eta\equiv|\eta_1-\eta_2|\ge0$ for the events in the
      rectangle region $\Delta\eta<b$ and
      $\eta^{}_B\equiv|\eta_1+\eta_2|<c$.
      The cut $\eta_B<c$ is necessary to reduce the effect of
      the parton distribution functions on different $\Delta\eta$ bins.
      The corresponding centrality ratio
      $R_{\Delta\eta}$ is defined as
\begin{equation}\label{deltaEtaR}
    R_{\Delta\eta}(a,b,c)=
\left.\frac{N(\Delta\eta<a)}{N(a<\Delta\eta<b)}\right|_{\eta^{}_B<c}.
\end{equation}
       The normalized histograms of the $\Delta\eta$-spectra and the
       theoretical curves are shown in
Fig.~\ref{fig:eta1} for the following
       parameters values: $b=3.5$ and $c=1.5$.
       It can be seen from the figure that the theoretical distributions
       describe very well the above-mentioned simulation data.
       In the same
       way as was done for the $\chi$-distribution of the excited bosons
       one can maximize the deviation of
       $R_{\Delta\eta}(a,b,c)$ from the QCD ratio
       $R^{\rm QCD}_{\Delta\eta}=N_{\rm QCD}^{(a)}/N_{\rm QCD}^{(b-a)}$.
       One finds the desired  minimum of
       $R_{\Delta\eta}(a,b,c)$, which corresponds to the maximal
       deviation from QCD, for $c=1.5$, $b=1$ and $a=0.67$
 \cite{Chizhov:2010jg}.

\section{Conclusion}

     In this     review     phenomenological consequences of the Standard Model
     extension by means of new spin-1 chiral fields with the internal quantum
     numbers of the electroweak Higgs doublets are     summarized.
     It is worth stressing that the new type of spin-1 chiral bosons can exist.
     They are well motivated from the 
     point of view of the Hierarchy problem and are predicted
     by at least three different classes of theories that represent
     different approaches to explanation of the relative lightness of the
     Higgs doublets.
     The decay distributions of the chiral bosons differ
     drastically from the distributions of the known gauge bosons
     and can be distinguished from others.
     The discovery of this type of distributions will point
     to existence of compositeness, new symmetry and even extra dimensions.

     The prospects for resonance production and detection
     of the chiral vector $Z^*$ and $W^{*\pm}$ bosons
     at the LHC energies are considered on the basis of
     quantitative simulations 
     within the CompHEP/CalcHEP package.
     The experimental signatures of the
     excited chiral heavy $Z^*$  bosons 
     are considered and compared with those of the gauge $Z'$ bosons.

      The $Z^*$ boson can be observed as a Breit-Wigner resonance peak
      in the invariant dilepton mass distributions
      in the same way as the well-known extra gauge $Z'$ bosons.
      This naturally puts the chiral bosons on the list of
      very interesting objects for early searches with the first LHC data.
      Moreover, $Z^*$ bosons have unique signatures in transverse momentum,
      angular and pseudorapidity distributions of the final
      decay products, which allow one to distinguish
      them from other heavy neutral resonances.
      In particular, there is no Jacobian peak in the
      transverse momentum distribution of the decay products,
      and the  angular distribution (in the Collins-Soper frame for high on-peak
      invariant masses) has a peculiar       ``swallowtail'' shape.

      In 2010, with 40~pb$^{-1}$ of the LHC proton-proton data at 
      energy 7~TeV, the ATLAS detector was used to search for narrow resonances
      in the invariant mass spectrum 
      of $e^+e^-$ and $\mu^+\mu^-$
      final states and high-mass charged states
      decaying to a charged lepton and a neutrino.
      No statistically significant excess above the Standard Model expectation
      was observed.
      Therefore, low mass limits of 1.15~TeV$/c^2$ and 1.35~TeV$/c^2$
      were obtained for the neutral chiral $Z^*$ and charged $W^*$ bosons,
      respectively.
      These are the first direct
      limits on the $W^*$ and $Z^*$ boson production.

      Finally,
      a novel strategy for the neutral chiral boson search 
      in the LHC dijet data       is discussed.
      For 
      almost all currently considered exotic models the 
      relevant signal is expected in
      the central dijet rapidity region $y_{1,2}\simeq 0$ and $|y_1-y_2|\simeq 0$.
      Contrariwise,       the excited bosons
      do not contribute to this region
      but produce an excess of dijet events
      far away from it. 
      In particular, for these bosons the appropriate
      kinematic restrictions
      can lead to a dip in the centrality ratio distribution over the
      dijet invariant mass instead of a bump expected in the most exotic  models.

      We expect that the experimental results, presented here and based on low statistics
      of 2010 will be very soon improved with much higher statistics of 2011.
      Furthermore, we plan to extend our search for excited bosons by means of
      thorough investigation of the dijet data from the ATLAS detector at the LHC.

\smallskip

\section*{Acknowledgments}
     We are very grateful to O. Fedin and his colleagues
     for fruitful cooperation, V.G. Kadyshevsky and N.A. Russakovich
     for support and interest in this research.
     The work of M.V. Chizhov was partially supported by the grant of
     the Plenipotentiary of the Republic of Bulgaria in JINR for the year 2011.

\providecommand{\href}[2]{#2}\begingroup\raggedright\endgroup

\end{document}